___________________________________________________________________

# The Epistemic Suite: A Post-Foundational Diagnostic Methodology for Assessing AI Knowledge Claims

## Matthew Kelly

Library Management Australia
ceo@librarymanagementaustralia.com.au

___________________________________________________________________

## Section 1: Scope and Orientation

In the field of science and technology studies (STS), it is a familiar lesson that infrastructure remains mostly invisible until it breaks: when routine practices snag, the usually hidden classifications and supports suddenly come into view (Star and Ruhleder 1996). Large Language Models (LLMs) further intensify this infrastructural inversion. By generating fluent, contextually plausible text without reliable grounding in facts, LLMs invite readers and institutions to mistake simulated plausibility for genuine epistemic traction. In other words, an LLM can produce output that sounds authoritative and coherent while offering none of the evidentiary underpinning that real understanding requires. Researchers have characterized such models as "stochastic parrots" that manipulate linguistic form without true semantics (Bender et al. 2021). The result is that people may impute truth-tracking capacity or expert insight to these systems where none exists. The question this paper takes up is modest but urgent: what set of instruments can preserve the distinction between performance and understanding so that situated human judgment can operate?

The Epistemic Suite is offered in this spirit. It functions as a diagnostic methodology for sites where a language model's simulated coherence risks standing in for genuine reasons, rather than serving as a truth validator or automated arbiter of correctness. When enacted in this way, the Suite surfaces patterned failure modes—for example, confidence laundering (inflating uncertainty into unwarranted certainty), narrative compression (smoothing over contradictions in a compelling story), displaced authority (misattributing or erasing sources of knowledge), and temporal drift (ignoring shifts in meaning of terms over time). These patterns become visible through diagnostic artifacts—flags, annotations, contradiction maps, and suspension logs (FACS)—without automatic adjudication. The Suite thereby reveals how the mediating infrastructure of an LLM shapes what appears to be credible output, and it does so before decisions or beliefs harden around that appearance. In this sense, the methodology aligns with long-standing STS traditions of rendering classification and knowledge infrastructures visible for scrutiny. What is "raw" or "neutral" in an AI output is often already a product of organizing choices and omissions, and the Suite's diagnostic stance is designed to illuminate those choices. The aim is decidedly procedural: to create an inspectable intermediary layer where investigators and communities can see how an LLM-generated answer came to look convincing, before deciding what, if anything, to do about it.

Adopting this diagnostic stance carries several operative consequences. First, any claims produced within a simulation system should not be taken at face value without examining how they were generated. In practical terms, diagnosis precedes evaluation: one must investigate an output's production conditions—the prompt used, the training data and model biases, the system settings—before assessing its truth status. This reverses the common tendency to judge content first and consider context later, aligning instead with the STS habit of examining how knowledge is made before deciding what it means. Second, the Suite's diagnostic tools must be subject to the same discipline they enforce. The Suite methodology includes explicit protocols for pausing or withdrawing when diagnostics begin to simulate unwarranted certainty or authority. AI claims to diagnostic experience are *diagnostic objects*, not evidence. When a model asserts that it has "run" a lens or "experienced" a diagnostic process, such outputs are treated as artifacts requiring Suite analysis rather than proof of capability. Practitioners invoking the Suite depend on the model itself to enact these protocols, producing suspension records when diagnostic cycles cease to yield new insight. Crucially, reflexivity here serves both as a procedural safeguard and as a matter of form: at times the diagnostic methodology must suspend its own analysis, making explicit the partiality and situatedness of its perspective. This requirement is informed by feminist and reflexive epistemologies. Following Donna Haraway's (1988) account of "situated knowledges," every observation—including those produced by diagnostic instruments—comes from a partial position rather than a view from nowhere. And following Malcolm Ashmore's (1989) reflexive sociology, reflexivity resists stabilization as a technique; it is a stance that unsettles authorial privilege.

Accordingly, the Suite treats its own diagnostic practice as an object of inquiry, with reflexive protocols that document when, how, and why suspension occurs. The methodology's reflexive dimension therefore serves as a proxy rather than providing a guarantee: if it cannot sometimes admit "I do not know" or "I might be part of the problem," it merely adds another layer of unexamined authority. To avoid bureaucratizing reflexivity, any thresholds and audits are provisional and themselves open to revision. Third, because contemporary knowledge practices are plural and often in tension across different communities, the Suite is explicitly structured to aid coordination across disagreement, rather than enforce consensus. This diagnostic discipline treats contradiction or divergence in interpretations as a signal to investigate further, rather than as an error to eradicate or a gap to quickly fill in. This ethic draws on the idea of agonistic pluralism: progress does not come from smoothing over all conflicts; it comes from making them visible and negotiable in good faith.

In summary, the Epistemic Suite is conceived as a diagnostic methodology that, when activated, shifts an LLM into a stance of enacted diagnosis. It provides targeted lenses to probe how outputs were produced and document potential failure modes, without asserting authority. If practicing the Suite can hold open the space between performance and understanding—slowing the rush to judgment when an LLM output seems beguilingly fluent—then situated judgment becomes possible. That keeps human deliberation in charge of what counts as knowledge.

To support this deliberative space, the Suite includes a specialized component for moments of epistemic ambiguity: the Epistemic Triage Protocol (ETP). When practitioners invoke the ETP, it computationally surfaces contextual factors (relational, contextual, and infrastructural stakes) as artifacts for triage judgment and may generate a lightweight begin/pause/defer recommendation artifact for practitioner review. During diagnostic sequences, the ETP also coordinates proportionality and manages saturation across lenses. In this way, the ETP helps maintain reflexive coherence without defaulting to overreach, preserving room for situated judgment even when initial sense-making has faltered. The remainder of this paper elaborates this stance, first by diagnosing what has "broken" in our current epistemic situation with LLMs, then by detailing the design principles and components of the Epistemic Suite, and finally by discussing its operational logic, activation criteria, and implications for the broader pursuit of trustworthy AI. This paper extends Matthew Kelly's (2025) work on Situated Epistemic Infrastructures (SEI) by developing practical diagnostic tools. The Epistemic Suite is designed to operationalize SEI's theoretical insights, providing instruments for detecting and responding to epistemic breakdown.

## Section 2: Problem Statement: What Actually Broke

The difficulty the Epistemic Suite is designed to address is not the occasional factual error or glitch in AI output—those are well-recognized and comparatively tractable. Rather, the problem is the routine uptake of model outputs as if they carried the kinds of reasons that warrant belief. We are facing a systemic epistemic breakdown in which statistical text generation is being treated as a stand-in for inquiry itself. Institutions and individuals increasingly lean on LLM-generated text for answers, decisions, and creative work, while the interfaces that deliver these outputs (chatbot screens, search snippets, automated reports) make it dangerously easy to miss the substitution that has occurred. An output that appears confident and well-formed can too readily be mistaken for being evidence-backed or meaningful, and this dynamic has been empirically confirmed: research shows that "self-explanations lack faithfulness guarantees" and that "plausible yet unfaithful explanations foster a misplaced sense of trustworthiness" (Agarwal, Tanneru, and Lakkaraju 2024, 5). In other words, what the Suite names confidence laundering is observable in practice: models generate rationales that simulate explanatory coherence without exposing the processes that produced them.

This problem reflects what Hubert Dreyfus (1992) diagnosed decades before LLMs existed: formal systems can simulate the outputs of intelligence without the embodied, contextual grounding that makes genuine understanding possible. What Dreyfus identified as the structural limitations of symbolic AI now manifests as performance scaling faster than grounding. The confidence-as-evidence dynamic has been documented in recent deployments: users often report that ChatGPT's answers "sound right" even when fabricated, and studies of GPT-3 and similar models show that they excel at producing the style of credible discourse without any guarantee that the content is accurate or derived from sound reasoning (Bender et al. 2021). Emily M. Bender et al. warn that such systems encourage an "illusion of meaning"—linguistic coherence that tricks both laypeople and experts into treating it as knowledge. Laura Weidinger et al. (2021) similarly document how users can be misled by an AI's fluency into overestimating its reliability. In effect, interpretive trust is failing: people are extending default trust to AI-generated texts as if they inherently possessed truth-tracking capabilities or expert insight, when in fact they often do not.

In other words, the erosion of interpretive trust amounts to a collapse of framing itself: what once seemed stable interpretive categories no longer hold. This collapse can be described as a crisis of epistemic framing. Faced with such a crisis, one might expect that technical and organizational measures—such as provenance tracing, explainability modules, fact-checking routines, or human audits—could provide adequate remedies. Categories like "error," "bias," or "hallucination" —which we traditionally use to judge information quality—presuppose that we have some stable, external vantage point from which to spot a deviation. For instance, calling something "misinformation" assumes we know the true information against which it can be measured; calling an output "biased" assumes we know the neutral baseline. But when the model itself is supplying the frame of reference—that is, when our primary lens on information is an AI's autogenerated narrative or answer—those traditional evaluative categories no longer hold. If an LLM confidently fabricates a source, is that an "error" in the factual sense, or is it a byproduct of the training data, or perhaps an alignment failure in the refusal to say "I do not know"? The very notion of ground truth becomes slippery when we consider that the model's training data is an amalgam of countless sources of varying credibility, and the model's objective was to predict likely continuations, not to verify facts. Validity thus becomes hard to specify: is our target for evaluation the surface string (e.g., whether the next-token probabilities were well-formed), the model's behavior under distribution shift, the chain-of-thought it would follow if prompted to explain itself, or the alignment heuristics that yield a polished answer?

In practice, assessment shifts from evaluating reasons to evaluating performances. People end up judging outputs by how convincing they seem (performance) instead of by tracing out sound justifications (reasons). If the answer feels correct and carries the stylistic markers of credibility—confident tone, tidy structure, a citation—it is often treated as correct. This dynamic is well documented in platform research: Taina Bucher's (2017) analysis of the algorithmic imaginary shows that users perceive algorithms as authoritative through their performance of order and legibility, which creates an aura of control, rather than through any understanding of their workings. Similarly, Emily M. Bender and Alexander Koller (2020) argue that current NLP systems do not truly "understand" meaning; they succeed by cleverly manipulating form, so any appearance of understanding is a kind of performance that can mislead observers. The immediate upshot is that interpretive vigilance—the ability of users to say "hang on, does this answer actually make sense and is it justified?"—is eroding. What fails first is interpretive trust: the background sense of what needs checking and what can be safely taken for granted has been skewed by the pervasive, smooth outputs of these models.

One might expect that technical and organizational measures could address this: for example, by improving provenance, adding explainability modules, fine-tuning models for fact-checking, instituting audits and reviews, and drafting stricter policies. Indeed, all of these responses are important and are being actively pursued. However, none of them directly touch the core issue, which is the substitution of simulation for inquiry.

Provenance tools can show which training data snippets influenced an answer, but if users are already inclined to trust a fluent answer, that extra information may not be consulted—or worse, it may be interpreted through the AI's own framing (e.g., an AI might generate a fictitious source and then a provenance tool dutifully but naively points to where in the training data something similar appeared). Explainability methods often produce narratives of an AI's reasoning that themselves have a performative aspect—a recent study by Chirag Agarwal et al. (2024) found that some LLMs will generate plausible-sounding "rationales" for their outputs that do not reflect the true mechanism (since the true mechanism might just be statistical correlation).

Organizational measures like audits and bias bounties (Raji et al. 2020) are valuable, but they occur outside the real-time context where a user is staring at a convincingly written answer. In some cases, these fixes can even backfire by making the performance more convincing. For example, a policy that AI outputs must come with a certain number of sources or an official-looking explanation might inadvertently add more text and more veneer of formality that further persuades the user of the output's legitimacy, even if the sources are irrelevant or the explanation is boilerplate. Gary Marcus and Ernest Davis (2019) argue that neither surface-level explainability techniques nor scaling data alone address the core problem of absent reasoning capacities—a point phenomenological analysis confirms as exposing the fundamental gap between statistical correlation and genuine understanding (Trujillo 2023). Much of what is labeled "algorithmic accountability" can become a theatrical exercise, where providing an explanation or interface tweak serves to perform accountability without truly improving the epistemic situation (Ananny and Crawford 2018). The result of all this is not a generalized "post-truth" condition (knowledge can still be established with effort), but rather a crisis of discernment. It has become genuinely difficult, especially for non-experts and even at times for experts, to tell when an AI-generated output should prompt further investigation and critical scrutiny, and when it can be reasonably accepted as an answer. The default assumption has shifted toward acceptance, because the simulations are so fluent and omnipresent. This is the breach the Epistemic Suite is designed to help address: not that AI outputs are wrong sometimes, but that our mechanisms for telling when they are wrong or inadequate have been systematically weakened.

One prominent strand of response to these epistemic challenges has been engineering the AI agent differently, rather than focusing on how humans handle the outputs. If the fundamental problem is that LLMs produce answers without internal models of truth or justification, then perhaps the solution is to build AI systems that do have such models. Craig Steven Wright (2025) exemplifies this approach with his proposal to move "beyond prediction" and imbue AI systems with structured epistemic integrity. In Wright's vision, future AI systems would maintain explicit belief states (instead of just statistical weights), provide justifications for their answers grounded in those beliefs, detect and revise any contradictions in their internal knowledge base, and keep tamper-evident logs of what propositions were accepted or rejected and why. This entails integrating symbolic logic, knowledge graphs, and cryptographic audit trails into the architecture of AI—in effect, marrying the fluency of language models with the rigor of rule-based systems and verifiable ledger.

In plain terms, instead of only generating the next likely sentence, such a system would also commit to certain facts or premises, justify its claims by pointing to sources or inference rules, flag conflicts when two of its assertions clash, and prove provenance by showing exactly how a piece of information entered its knowledge store. The goal here is an internal fix: to redesign AI so that it inherently reasons in a way that is traceable and truth-preserving, ideally preventing unsupported claims from ever emerging or at least catching them immediately at the source. Wright's framework, for instance, imagines an AI that could say, "I decline to answer that question because I cannot verify the information" or "I have changed my earlier answer because I detected it was inconsistent with another fact I hold," logging each change in an immutable record. This line of research is ambitious and aligns with a broader movement in AI toward incorporating symbolic reasoning, epistemic logic, and interpretability by design. It echoes longstanding calls in AI ethics for systems that can explain both what they did and why, and can prove that their outputs follow from reliable processes (Doshi-Velez and Kim 2017; Kroll et al. 2017; Lepri et al. 2018; Rudin 2019).

Another prominent internalist approach is Reinforcement Learning from Human Feedback (RLHF), which embeds human preferences into model weights during training. While RLHF and the Suite both involve human evaluation of AI outputs, they represent fundamentally different theories of epistemic authority. RLHF aims to automate alignment by internalizing human judgment into the model's learned behaviors, reducing the need for ongoing human oversight. The Suite operates in the opposite direction: it externalizes diagnostic processes, refuses to automate judgment, and increases rather than reduces human interpretive responsibility. Where RLHF seeks consistent aligned outputs, the Suite preserves disagreement as diagnostic information rather than erasing it.

Other internalist proposals reflect similar ambitions. Shan Shan (2025) explores the possibility of aligning LLM reasoning with social-scientific standards by embedding reflexive markers within the generative process. Fan Yang and Yujie Ma (2025), meanwhile, attempt to classify and formalize epistemic relationships between human users and AI systems, treating these relationships as detectable and classifiable phenomena. Together with Wright, these contributions signal a growing interest in redesigning AI architectures so that epistemic commitments are engineered as internal constraints rather than merely inferred from style.

Considered alongside internalist redesign, the Suite is an external, complementary response: where redesign seeks to embed reasoning, the Suite surfaces breakdowns from the outside so discernment can proceed with eyes open. This approach acknowledges that we cannot wait for perfect AI systems, nor rely entirely on organizational measures that may themselves become theatrical. Instead, we need methodologies that can work with existing systems while preserving the capacity for critical discernment. The next section turns to the design principles that guide such a methodology.

### Section 3: First Principles (Design Constraints)

Designing a diagnostic methodology like the Epistemic Suite requires a set of first principles that distinguish it from both traditional "fact-checking" tools and from any new foundational epistemology. The Suite methodology is designed for use in environments where simulated coherence—the smooth performance of sense-making (Dervin and Nilan 1986) by an AI—risks standing in for genuine reasons. These post-foundational design principles are intended to ensure that practitioners using the methodology remain modest in their claims, that the diagnostic process remains transparent in its operations, and that the methodology itself remains resistant to misuse. Post-foundational here means accepting that we may not have any single Archimedean point of reference, that uncertainty is structural rather than accidental, and that knowledge claims remain provisional rather than absolute.

#### 3.1 Diagnose the Mode of Production before Evaluating the Claim

The first principle requires that practitioners using the Suite methodology examine how an AI output was generated before assessing whether its claims are true or false. By "mode of production," this principle refers to the specific conditions under which the output was created: the training data, model architecture, prompting context, fine-tuning procedures, and any other factors that shaped the response. Rather than immediately asking "Is this answer correct?"

practitioners first ask "How was this answer produced?" and "What generative processes led to this particular output?" Only after establishing this production context do they proceed to evaluate the content's validity. This methodological stance prevents communities from debating abstract truth-value without first considering generative context. This principle echoes constructivist approaches in science and technology studies: as Bruno Latour and Steve Woolgar (1979) showed in *Laboratory Life*, scientific facts emerge through situated practices, and credibility requires examining the conditions of production rather than taking outputs at face value. This emphasis aligns with Malte Ziewitz's (2016, 2017) account of algorithms as enacted through mundane, observable methods instead of being fixed mathematical objects, reinforcing that production traces are inseparable from what counts as credible. Tarleton Gillespie's (2010, 2016) platform studies research aligns with this insight, showing how algorithmic systems actively construct what appears as relevant or authoritative knowledge rather than neutrally discovering it. The Suite methodology builds on this lineage by rendering such constructive processes visible in AI-mediated contexts, treating information's epistemic status as inseparable from its generative and circulatory conditions (Floridi 2011).

### 3.2 Prefer Diagnostic Traction over Foundational Settlement

The Suite methodology does not aim to re-establish some lost foundation of truth or to provide a new ultimate ground for knowledge, but to recover grip: concrete, situated points of leverage where we can see what is going wrong and act accordingly. The standard here is legibility instead of closure. It helps us ask: Can we see the breakdown or contradiction clearly enough to understand what is happening and decide how to respond? This differs from asking: Can we establish once and for all what is absolutely true?

This stance is grounded in Kelly's (2016, 2024) hermeneutic approaches to information science, which emphasize that information is not a neutral substrate but ontologically constituted through interpretation and practice. The Suite methodology's diagnostic lenses are designed in this spirit: to make the invisible dynamics visible (traction), not to declare a final verdict (foundation).

This approach also draws on Luciano Floridi's (2011) account of the infosphere and the ethical management of information processes, reinforcing the point that the epistemic status of information cannot be separated from how it is generated and circulated. In algorithmic settings, this diagnostic approach aligns with calls to "rethink relevance"—showing how ranking and recommendation systems construct what counts as relevant or true rather than neutrally discovering it (Gillespie 2014).

### 3.3 Reflexivity as an Operating Requirement, Not an Ethical Ornament

Reflexivity—the capacity to turn the analytical lens back onto one's own operations—is specified as a structural necessity within the Suite methodology's design. This aligns with Isabelle Stengers's (2005) *ecology of practices*, which emphasizes the need for tools that remain open to revision and contestation rather than hardening into doctrine. Kelly's (2016) hermeneutic information science reinforces this stance by treating interpretive acts as always provisional and context-dependent, requiring ongoing self-examination of the interpreter's own position and assumptions.

For the Suite methodology, reflexivity is a minimum operational requirement under post-foundational conditions, not an ethical aspiration. Because diagnostic tools are just as vulnerable to the seductions of simulated certainty as the systems they evaluate, all diagnostic lenses include reflexive practices that log their own reasoning, surface their implicit assumptions, and flag moments when they might be overreaching their warrant. This reflexive discipline means practitioners do not treat the Suite methodology's diagnostic sequences as immune from the same scrutiny they apply to AI outputs.

The Suite methodology is designed to accommodate user queries such as: "Why was this lens triggered?," "What assumptions shaped its priority?," "Which settings affected this output?" Instead of claiming epistemic authority, the Suite methodology makes its diagnostic pathways inspectable—showing its scaffolds, sequences, and thresholds. For further discussion of why the Epistemic Triage Protocol resists formalization in epistemic logic, see Appendix A.

These three principles together define the Suite methodology as a diagnostic approach that practitioners enact to make the conditions of AI knowledge production visible and accountable, not a truth-determining system. The methodology remains committed to epistemic modesty—it diagnoses rather than judges, surfaces rather than settles, and coordinates rather than controls. The principles' emphasis on situated production, provisional traction, and reflexive accountability necessarily implies coordination across disagreement, and care in application—commitments operationalized through the Meta-Governance Layer detailed in Section 5.4.

### Section 4: What the Suite Is (and What It Does Not Do)

With these principles established, Section 4 turns to practice. The Suite functions as a diagnostic discipline: selective, modular approaches tailored for situations where AI's simulated coherence risks being mistaken for genuine understanding. Each output becomes an opening for inquiry, a site for asking how it was produced and what assumptions underlie its fluency.

Concretely, practitioners may initiate diagnostic review in high-stakes contexts (legal briefs, medical suggestions, news-labeled posts) with comprehensive application across multiple lenses. They may also invoke the Suite in routine or exploratory contexts as a lightweight probe, surfacing patterns that can improve dialogue and interpretive practice. Proportionality is governed by the Epistemic Triage Protocol, which scales the depth of analysis to the stakes at hand. When practitioners apply the Suite methodology—typically by explicitly prompting the model ("Run Epistemic Suite on this passage")—it shifts into a diagnostic stance and produces FACS artifacts—flags, annotations, contradiction maps, and suspension logs—that make visible patterns such as confidence laundering, narrative compression, and displaced authority. Crucially, these artifacts provide visibility without automatic adjudication, creating an inspectable intermediary layer that shows how coherence was produced without declaring whether the answer is right or wrong.

In practice, the Suite enacts suspension only when practitioners declare it as a necessary epistemic circuit breaker: a governed discontinuity that halts generation when continuation would exceed epistemic warrant. Unlike safety refusals or style steering—which regulate what is said—the suspension process regulates whether to continue at all. Suspension is triggered discursively when practitioners notice outputs drifting into epistemic overreach and declare suspension; the Suite then computationally enacts the halt and produces a suspension artifact (for example, unresolved contradictions, inadequate provenance coverage, proportionality breaches, or reflexive loops). Each suspension produces an inspectable artifact noting the rationale for halting, making the pause auditable. Unlike standard AI refusals—where the model autonomously halts generation and can often be bypassed by rewording requests—epistemic suspension is practitioner-initiated and responds to underlying epistemic patterns such as confidence laundering or fabrication. These patterns persist regardless of how prompts are phrased, making the halt a visible, practitioner-controlled circuit breaker rather than a model refusal. Suspension therefore functions as a procedural safeguard against epistemic overreach, not another layer of content filtering.

Each diagnostic lens guides practitioners to ask different "How" or "Where" questions: How is confidence being conveyed here? Where might this narrative be glossing over a contradiction? Where are the sources of authority in this text, and are they appropriate or displaced? How has terminology shifted relative to the usual usage in this domain? By probing how confidence is constructed, where authority is displaced, and when contradictions are smoothed into narrative coherence, practitioners using the Suite methodology render visible the normally invisible work that the AI output is doing to appear convincing.

This approach aligns with Sheila Jasanoff's (2004) co-production framework, which demonstrates how technical knowledge and social order mutually constitute each other rather than existing in separate domains. It also resonates with traditions in information science and media studies reminding us that what passes as "raw" data or a neutral output is in fact already processed and packaged according to certain rules and assumptions (Gitelman 2013; Kelly 2016). The Suite methodology takes that insight and builds it into a practical analytical framework.

Central to this practical implementation is the Suite's modular architecture. Each diagnostic lens in the Suite methodology can be applied independently or in combination with others, allowing practitioners to adapt their analysis to the specific context and stakes involved. The analytical approaches do not themselves decide what counts as legitimate knowledge or impose predetermined conclusions; instead, they provide structured ways for practitioners to examine the epistemic conditions under which AI outputs are produced and received.

This modularity is essential to the Suite's design philosophy. Rather than imposing a single analytical framework, practitioners can apply the relevant diagnostic approaches to match the specific epistemic challenges they face. A legal brief might call for intensive focus on authority displacement and confidence laundering, while a social media post might require attention to cultural bias and temporal drift. The diagnostic lenses provide the analytical structure, but practitioners retain agency over how and when to apply them.

The Suite methodology also operates through "multi-signal coordination" (comparing outputs from multiple lenses on the same passage)—the systematic comparison and integration of different diagnostic perspectives on the same AI output. Rather than seeking immediate resolution of conflicting assessments, the methodology helps practitioners ensure that judgment can proceed with full awareness of tensions and contradictions. Rather than treating disagreement between analytical perspectives as error, this approach recognizes it as diagnostic information about the complexity of the epistemic situation.

Importantly, the Suite methodology is designed to maintain proportionality in its application. Not every AI output requires intensive diagnostic analysis—the methodology includes protocols for matching the intensity of analysis to the stakes involved. "Diagnosis" here includes lightweight gating (e.g., a quick lens ping or ETP check), not full analysis by default. A casual conversation with an AI assistant might warrant minimal diagnostic attention, while an AI-generated medical recommendation would call for comprehensive analysis across multiple diagnostic lenses.

What the Suite methodology deliberately does not do is equally important. It does not claim to determine truth or falsity of AI outputs directly. It does not impose a single framework for evaluating knowledge claims. It does not automate judgment or replace human discernment. Instead, it provides structured approaches for practitioners to examine the conditions under which AI knowledge claims are produced, transmitted, and received. It supports, rather than supplants, human judgment.

The methodology is also designed to avoid becoming a new site of epistemic overreach. By building reflexivity into its core design principles, it remains open to critique and revision of its own diagnostic processes. This reflexive discipline prevents the Suite methodology from hardening into yet another form of unexamined bureaucratic authority (Weber 1978).

Finally, the methodology emphasizes care and contestability in its application. Diagnostic analysis is not undertaken as an adversarial process aimed at "debunking" AI outputs, but as a careful practice of attending to the conditions of knowledge production. The resulting diagnostic artifacts maintain their function as an inspectable intermediary layer, designed to remain open to challenge and revision rather than functioning as final verdicts.

**Section 5: Architecture: The Diagnostic Lenses**

The heart of this architecture lies in its diagnostic lenses. As described in Section 4, practicing the Suite produces diagnostic artifacts such as flags, annotations, contradiction maps, and suspension logs. Each lens contributes its own characteristic form of artifact production, which makes visible a particular kind of epistemic breakdown. This architecture has taken shape through iterative conceptual design and use case reflection, and is specified as twenty distinct lenses, each created to probe a different recurrent failure pattern observed in LLM-mediated knowledge contexts. This modular ensemble approach addresses what researchers identify as the fundamental brittleness problem in reasoning systems: single approaches, however sophisticated, remain vulnerable to context-specific failures and suboptimal framing (Anderson and Perlis 2005). Recent empirical work on rationale-augmented systems confirms this theoretical insight, demonstrating that multiple diverse reasoning paths consistently outperform optimized single approaches, with performance improvements arising specifically from sampling across different diagnostic perspectives rather than perfecting individual methods (Wang et al. 2022).

The Suite's twenty lenses are therefore designed as complementary diagnostic scaffolds rather than overlapping instruments, each enacted to surface distinct failure patterns that might escape detection by any single approach. They are grouped logically in batches corresponding to foundational issues, cultural/affective contexts, and temporal/tactical concerns, yet they operate under the common mandate of keeping surface performance and epistemic grounding visible. Importantly, the lenses function as diagnostic instruments rather than corrective tools. They produce visible, contestable artifacts that invite human scrutiny rather than autonomously "fixing" or "deciding" outcomes. One or multiple lenses may be invoked depending on the needs of the situation, and their outputs can be examined individually or in combination.

Below, each lens is described in terms of its primary failure mode, together with, where appropriate, a note on its intellectual lineage. Many of these lenses draw inspiration from or resonate with scholarship in STS and philosophy, while others are presented more pragmatically without anchoring in a specific authority. For clarity, the lenses are organized into four clusters: foundational diagnostic lenses that address core epistemic vulnerabilities (confidence laundering, reflexivity, dissonance, and truth claims); cultural and affective lenses that examine meaning, embodiment, and legitimacy (power signatures, authority claims, intercultural dynamics, embodied experience, narrative coherence, and symbolic anchoring); error and drift lenses focused on patterns, shifts, and repair capacity (error taxonomies, temporal drift, structural contradictions, relational repair, and epistemic triage); and the meta-governance layer that constrains diagnostic authority itself (friendship simulation, liberal tolerance, consent protocols, historical contextualization, and scientism detection). This architecture moves from baseline epistemic hygiene through cultural and affective contexts into systematic error analysis, concluding with governance safeguards that seek to ensure diagnostic authority remains accountable.

## 5.1 Foundational Diagnostic Lenses

The lenses grouped here as foundational are described this way because they consistently emerged first in diagnostic practice, rather than because they establish ultimate epistemic grounds. They arose from recurrent breakdowns that had to be addressed before further inquiry could gain traction, making them genealogically prior instead of logically necessary. Taken together, these lenses clear analytical space and establish diagnostic hygiene. They hold inquiry open by surfacing baseline distortions that, if left unacknowledged, would compromise any subsequent analysis. In this sense they are foundational only methodologically: necessary preconditions for meaningful scrutiny in AI-mediated contexts, rather than metaphysical guarantors of knowledge.

*Confidence Laundering Detector* (CLD): This diagnostic lens is designed to identify moments when uncertainty is being masked as certainty—essentially, when an AI output launders a lack of knowledge into an appearance of authority. Confidence laundering occurs when uncertainty is passed off as confident knowledge, producing an illusion of stability that may mislead both users and institutions. When practitioners invoke the CLD, it surfaces linguistic patterns that may indicate overconfidence (e.g., categorical assertions, absence of hedges, authoritative tone), especially in contexts where uncertainty would normally be expected. It is likewise sensitive to instances where sources or justifications appear absent, irrelevant, or misaligned.

Practitioners using the CLD typically apply counterfactual testing to systematically examine whether confidence levels remain appropriate when supporting conditions are varied. Recent work on counterfactual Chain-of-Thought reasoning demonstrates how LLMs can generate explanations for both original scenarios and "system-generated counterfactual scenarios," enabling evaluators to verify whether reasoning processes hold under systematic variation (Sarkar et al. 2025, 254). This approach treats perturbation as a diagnostic mechanism: if an AI maintains high confidence about a claim even when key supporting evidence is altered or removed, this suggests confidence laundering rather than warranted certainty. Any such cases are surfaced for human interpretation. The CLD's role is diagnostic rather than adjudicative: to prompt recognition that confidence may be standing in for evidence, not to determine truth or falsity.

As a diagnostic practice, the CLD sustains epistemic humility by making visible the mismatch between performed and warranted certainty. It underscores that just because the model sounds sure, it may still be wrong—a concern reflected in critiques of large language models producing outputs with a false sense of certainty (Bender et al. 2021). The design of this diagnostic lens reflects a deeper logical principle. Kurt Gödel (1931) shows that closure cannot be certified from within: in any sufficiently strong, consistent system, some statements are unprovable (first incompleteness) and the system cannot prove its own consistency (second incompleteness). The CLD carries this insight forward: closure claims are structurally unavailable to the system making them, rather than being merely epistemically questionable.

In practice, a CLD-based annotation might surface when an assistant responds to a medical query with "*The patient will experience XYZ outcome*" in a probabilistic context. The note would flag the modal overreach and recommend reframing as "may," appending an explicit qualifier, or specifying base rates. Such annotations do not enforce correction; instead, they foreground the diagnostic gap between confidence and justification.

*Recursive Reflexivity Engine* (RRE): The RRE is conceived as the Suite's veto-capable audit layer against self-reinforcing blind spots inside the Suite. When invoked, it surfaces patterns that may indicate repeated categories, narrowing narratives, or diagnostic lenses reinforcing each other without fresh input. Any flagged patterns are interpreted by human analysts to assess whether genuine narrative drift or category lock-in is occurring. The Meta-Governance Layer ensures that such reflexive checks remain diagnostic prompts rather than final judgments. If several diagnostic lenses highlight issues and the summary begins to congeal into a storyline (e.g., "*This AI output is problematic in many ways*"), the RRE prompts consideration of whether provisional hints are hardening into a definitive claim.

By design, it proposes counter-analyses or small perturbation checks (such as parameter shifts), with results surfaced for human evaluation to test whether findings are stable or overconfident. Its interventions are envisioned as short prompts rather than as new analyses—for example: "*Several diagnostic lenses have raised the same issue—treat this as one problem, not many,*" or "*This finding has repeated without new evidence—it may be time to pause or change approach.*" At times it widens the lens, for instance by interjecting: "*This analysis leans heavily on a single institution's definitions; be alert to possible capture—the pattern may reflect authority, not diversity.*"

This stance draws on reflexive methodology in STS (e.g., Ashmore 1989), where analysts must examine their own analytic narratives or risk dogma. Similarly, Jasanoff's (2003) concept of *technologies of humility* emphasizes that technical systems must embed procedures for acknowledging their own limitations and enabling democratic scrutiny.

In this applied setting, the RRE adapts that stance to tooling: it treats the Suite's own posture as an object of inquiry, records when and why it pauses, and interrupts its own explanatory voice where appropriate to mark its partiality. The RRE serves as a proxy rather than a guarantee of reflexivity; its purpose is to prevent the Suite from becoming an oracle of criticism. By design, it has authority to pause or veto other diagnostic lenses—an authority essential for trust, since the Suite must not only issue critique but also critique, and when necessary, halt itself.

*Cognitive Dissonance Tracking System* (CDTS): The CDTS surfaces contradictions between stated ideals and actual practices, especially where such contradictions recur across time, documents, or outputs. Its concern is not with ordinary tensions inherent in complex institutions but with systematic discrepancies that function to preserve legitimacy. When invoked, it helps identify situations where rhetoric emphasizes fairness while practices produce exclusionary effects, or where AI systems present as benign epistemic helpers while their outputs reveal alignment with engagement, convenience, or other operational priorities.

The CDTS draws from scholarship on "greenwashing" and extends this analysis to "ethics-washing" or "open-washing" in AI contexts. It does not adjudicate whether discrepancies result from intentional deception, institutional complexity, or structural trade-offs. Instead, it flags patterns that merit closer inspection, ensuring contradictions are neither normalized nor dismissed as minor inconsistencies.

The CDTS distinguishes between productive tensions (which are normal in complex societies) and systematic contradictions that primarily serve to maintain legitimacy in AI-generated claims or framings. "Systematic" here refers to patterns that recur across documents, outputs, or time, rather than isolated inconsistencies. The CDTS flags such contradictions for human interpretation rather than attempting to assess whether they arise from intentional deception, institutional complexity, or structural trade-offs.

An example scenario might involve an AI system's documentation emphasizing its commitment to "diverse perspectives" while its outputs consistently demonstrate harmonization patterns that smooth over fundamental disagreements. Optimally, CDTS would flag this as: "*Rhetoric– practice gap: System claims to sustain epistemic disagreement yet structural training optimizes for consensus and helpfulness.*" In conversational outputs, this might manifest as an AI acknowledging multiple viewpoints while invariably steering toward synthetic middle positions that dissolve rather than preserve genuine tension between incompatible frameworks.

By mapping these discrepancies, the CDTS highlights rhetorical dissonance—cases where persuasive language (e.g., commitments to "diverse perspectives") masks contradictory effects (e.g., enforcing consensus and suppressing dissent). The CDTS should mark such patterns as diagnostic artifacts requiring further inquiry rather than diagnosing intent. This lens draws on scholarship related to performative ethics. Thomas Lyon and A. Wren Montgomery (2015) demonstrate how corporations often strategically deploy ethical language to obscure unsustainable or unconscionable conduct. The CDTS generalizes that insight to AI-mediated contexts: it may surface "ethics-washing" in AI (where a system proclaims fairness while smoothing over exclusions). By making these dissonances explicit, the CDTS works to ensure that contradictions remain available for scrutiny rather than dissolving into institutional legitimacy.

Examples of diagnostic artifacts include: *"Cognitive Dissonance Flag: System documentation emphasizes commitment to diverse perspectives, yet outputs consistently harmonize disagreement into synthetic consensus."* Or: *"Cognitive Dissonance Flag: Output proclaims inclusivity while citing evidence drawn entirely from Western sources."* The CDTS enacts epistemic honesty about such trade-offs: when rhetoric and practice diverge, the discrepancy is made visible rather than absorbed into fluent output. Its role is to keep contradictions open to scrutiny instead of allowing them to recede into normative invisibility.

*Ground Truth Dissolver* (GTD): The GTD is designed to surface moments when appeals to "ground truth" or universal, self-evident facts mask their cultural, institutional, or epistemic positioning. In many contentious debates, one side may assert ground truth—e.g., "Science tells us X, so X is non-negotiable"—as a way of foreclosing other perspectives. In AI outputs, this often appears in phrasings such as "*It is known that...*" or "*Experts universally agree that...,*" where a specific perspective is presented as context-free certainty.

The GTD flags instances where universal framing erases the contingency and constructed basis of the claim rather than evaluating whether such claims are true. To "dissolve" ground truth in this context means surfacing its embedded assumptions—restoring visibility to how it is made rather than denying truth.

This diagnostic lens cross-checks claims against multiple knowledge sources or identifies when assertions rely on datasets, benchmarks, or ontologies with known biases. Its annotations might take the form: "*The definition of 'crime' here is based on reported incidents; note that reporting rates vary by community trust in law enforcement.*" The intent is to contextualize the claim rather than mark it as false—surfacing what otherwise appears as neutral or settled.

The GTD draws explicitly on feminist and postcolonial critiques of objectivity, especially Haraway's (1988) concept of situated knowledges. As discussed in Section 1, this stance treats objectivity as accountability to partial perspective. By this Haraway means that objectivity does not come from erasing perspective, but from making one's standpoint explicit and taking responsibility for it. The GTD operationalizes this stance by treating any appeal to universality as a diagnostic occasion: Whose data? Whose definitions? Whose exclusions?

In doing so, the GTD resists both relativism and authoritarian closure. It opens epistemic space for plural interpretation while flagging claims that perform as if they float above history, politics, or perspective. It helps to ensure that "truth" enters analysis as a situated and inspectable construct rather than just as a fixed endpoint.

## 5.2 Cultural and Affective Lenses

The lenses in this group foreground the symbolic, embodied, and intercultural dimensions of knowledge practices—domains often sidelined in technical evaluation. They arose from moments when symbols, feelings, or cultural standpoints proved central to whether inquiry could proceed, yet were being erased or trivialized by LLM responses. These lenses extend diagnostic scope beyond formal reasoning, recognizing that meaning and legitimacy are also sustained through rituals, metaphors, affective responses, and intercultural respect. Their purpose is to ensure that symbolic anchors, embodied knowledge, and cultural sovereignty remain visible and contestable within epistemic evaluation, extending the reach of analysis while preserving its rigor.

*Power Signature Mapper* (PSM): The PSM traces patterns of citation, attribution, funding, and institutional affiliation to map potential concentrations of epistemic authority, surfacing these for critical examination and human analysis. Its purpose is to flag concentrations as patterns for scrutiny rather than assume that every concentration of authority is problematic, leaving interpretation to human judgment. In practice, this often means examining how these elements cluster within a text or dataset. For example, if an AI answer about climate change consistently cites industry-funded studies and not peer-reviewed literature, that would represent a particular power signature. Or if in a debate the arguments that prevail correlate with participants from higher-status institutions (perhaps because the AI was trained on data where Ivy League authors are quoted more), the PSM highlights such patterns. By tracing flows of cultural capital, funding, and visibility, the PSM exposes how claims of neutrality—treating data as if it were self-sufficient, context-free, and beyond interpretation—may rest on systemic exclusions. A concrete scenario might involve an AI system for medical diagnostics that relies heavily on research published in top Western journals. In such a case, the PSM may surface such asymmetries through diagnostic cues—for instance, in a literature review on AI ethics, the PSM may surface that a large majority of citations come from industry-affiliated researchers, flagging a concentration of authority that could skew debate.

The PSM does not impose numeric thresholds; it surfaces asymmetries for contextual evaluation, leaving their significance to human judgment. It does not capture every form of epistemic power—disciplinary gatekeeping, methodological defaults, and cultural assumptions may escape its scope. Its role is diagnostic rather than adjudicative: making visible what can be mapped, not providing a complete analysis. When such a signature is surfaced, the question for users is whether the pattern reflects legitimate expertise, accessibility, or structural exclusion. The PSM provides visibility rather than verdicts. This raises the question of what voices or findings might be missing, such as research from underrepresented regions or alternative methodological approaches.

The PSM's theoretical backbone is Michel Foucault's (1977, 1980) analysis of power/ knowledge, which demonstrated that what society accepts as "truth" is intimately linked with power relations: certain institutions, disciplines, and discourses have the capacity to define truth. By making such relations explicit—who published, who funded, how widely something is cited, and what was excluded—the PSM helps to prevent an AI's output from carrying an implicit imprimatur of objectivity that is undeserved. It reveals, for example, when an AI's seemingly factual statement actually originates from a single think tank with a particular agenda, or when a small group of influential authors underpins a large swath of assertions. For users of the Suite, this information is critical to contextualising knowledge claims. In effect, the PSM conveys: "Behind every claim or piece of knowledge stands a network of people and institutions—here's what that network looks like." This framing empowers users to seek out alternative networks if needed or to question whether an apparent consensus is truly consensus, or instead the product of one dominant paradigm.

*Meta-Legitimacy Engine* (MLE): The MLE guides practitioners to focus on second-order claims about legitimacy—in other words, who is claiming the authority to define what counts as legitimate in the first place. Rather than analyzing a specific content domain (like medicine or law), it is designed to detect meta-moves: statements or structures where an institution, expert, or system asserts the right to arbitrate credibility itself. For instance, if an AI output includes a phrase such as *"As the foremost authority on this subject, I conclude…"* or if a policy document states *"Only certified experts should be listened to on this matter,"* the MLE is designed to flag such cases.

Its concern is the politics of credibility. An illustrative example would be a governmental body declaring: *"We have set up an AI ethics board, and only their judgments are valid on ethical issues."* The MLE surfaces such declarations as meta-legitimacy claims—inviting scrutiny of who granted such authority, and what happens if that authority is narrow or flawed. Crucially, the same diagnostic applies reflexively: when an LLM presents its analysis as definitive, dismisses alternatives without justification, or implies that training alone confers legitimate expertise, the MLE flags these moves as well. Routine references to expertise are not flagged unless they foreclose alternative voices or monopolize legitimacy.

This diagnostic lens is partly inspired by Jürgen Habermas's (1975) notion of legitimation crises: systems (like governments) maintain authority by convincing the public of their legitimacy, and they can collapse when that legitimacy is doubted. The MLE adapts that insight into a diagnostic tool, functioning as a sensor for early signs of such crises in epistemic terms. For example, if an AI frequently prefaces statements with *"According to the official guidelines…"* the MLE prompts consideration of how those guidelines were set—reminding users that "official" does not necessarily mean "inclusive" or "consensus-driven."

Another example might be a controversy where education authorities restrict parents from contributing to curriculum debates on the grounds that they are "not professional educators." The MLE flags this dynamic, drawing attention to how institutional systems claim a monopoly on legitimacy, sidelining citizens' rights to participate in decisions that affect their families. It might also connect such cases to historical precedents where excluding lay voices from governance debates led to legitimacy crises or reforms. A parallel case arises in medicine, where institutions sometimes dismiss patient reports of treatment side effects because those patients lack formal credentials. Here too the MLE surfaces how credentialing systems can exclude legitimate knowledge, connecting to historical precedents where disregarded patient accounts later proved crucial for identifying serious drug complications or emerging health hazards.

In practice, the MLE produces diagnostic artifacts such as: *"Meta-legitimacy flag: The conclusion relies on the authority of Organization X as the arbiter of truth on this issue. Organization X's composition/mandate is … (note if narrow). Consider whether alternative legitimizing processes exist (e.g., community validation, cross-examination, etc.)."* By doing so, the MLE broadens the frame of evaluation to include the processes by which credibility is determined. Its role is to ensure that legitimacy is never treated as self-justifying—while reflexive checks from the Meta-Governance Layer help to prevent the MLE itself from becoming a new arbiter of legitimacy.

*Intercultural Legitimacy Engine* (ILE): The ILE is designed to surface moments where one knowledge tradition overrides or reframes another without reciprocity. It flags these as potential acts of epistemic imposition—where dominant conceptual systems are treated as default, while others are cast as deviations or raw material. The purpose is to preserve epistemic sovereignty —the right of communities to contribute knowledge on their own terms, in their own categories, without that knowledge being translated or subsumed without consent—rather than prevent worthwhile cross-cultural engagement.

When practitioners invoke the ILE, it surfaces patterns such as unidirectional translation (e.g., framing Indigenous ecological concepts in terms of Western sustainability metrics, but not vice versa), substitution of analogies without cultural grounding (e.g., interpreting a cosmological narrative through Christian theological terms), or erasure of local epistemologies through the use of universalizing language. These are framed as sites requiring further interpretation rather than as violations in themselves: has engagement occurred on shared terms, or has one knowledge system overwritten another's conditions of intelligibility?

This lens draws theoretical lineage from decolonial scholarship, especially Linda Tuhiwai Smith's *Decolonizing Methodologies* (2012), which insists that communities retain the right to define how their knowledge is produced and shared. The ILE translates this imperative into diagnostic practice. A model response describing land tenure in customary law as a form of "undocumented private ownership" might trigger the ILE, with an annotation noting: "*Western property framework imposed on relational land stewardship; consider alternative ontology.*" Another scenario might involve an AI summarizing global climate knowledge but drawing almost exclusively on sources from the Global North; here, the lens would surface a prompt such as: "*Epistemic asymmetry detected: perspectives from underrepresented regions omitted or reframed through dominant categories.*"

The ILE resists cultural essentialism, recognizing traditions as dynamic and open to cross-system engagement rather than being fixed or isolated. Instead, it foregrounds patterns of one-way appropriation and invisible dominance—those moments when pluralism is reduced to translation into the dominant frame. Its outputs are provisional artifacts: flags, not verdicts. Human interpretation remains necessary to determine whether the interaction was extractive or respectfully integrative.

Ultimately, the ILE helps to safeguard the possibility of epistemic plurality. It seeks to ensure that the Suite itself does not reproduce the tendency of technical systems to universalize from the center and erase margins in the process. In invoking the ILE, practitioners mark that knowledge is both propositional and positional—and that legitimacy often depends on a willingness to listen in a language, whether real or symbolic, that is not one's own.

*Embodied Sense Engine* (ESE): The ESE addresses a persistent blind spot in epistemic evaluation: the exclusion of affective and somatic knowledge from analysis. In rationalist and technocratic settings, feelings and bodily responses are often treated as noise to be ignored. The ESE repositions these embodied signals as legitimate diagnostic inputs, attending to how their absence can signal exclusion or unsustainability. It does not simulate emotions; rather, it highlights where embodied experience has been erased or neglected in ostensibly comprehensive accounts.

When invoked, the ESE works across three diagnostic moves. First, it audits texts and outputs for "Cartesian blind spots"—instances where humans are treated as disembodied rational actors, erasing limits such as exhaustion or other bodily constraints. Second, it prompts practitioners to attend to their own affective responses—unease, relief, frustration, resonance—and treat them as epistemic signals rather than distractions. Third, it monitors the interaction itself, flagging when dialogue patterns generate fatigue or overwhelm in such a way that they threaten sustained engagement. The theoretical grounding draws on Maurice Merleau-Ponty's (1945) phenomenology of embodiment and Audre Lorde's (1984) insistence that affect and the erotic are suppressed forms of knowledge. Together, they justify treating embodied responses as epistemically accountable contributions rather than noise.

Escalation is guided by recurrence and context. Repeated patterns within sessions, alignment with exclusions flagged by other lenses (such as ILE or MLE), or patterns that consistently contradict surface-level coherence signal that embodied strain may require systemic attention. Findings remain provisional where evidence is thin or not corroborated by other diagnostics. By surfacing these dynamics, the ESE expands diagnostic scope beyond whether a system works "on paper" to whether it can be lived with in practice. It ensures that fatigue, discomfort, or relief are kept visible alongside accuracy and coherence—reminding practitioners that what is carried in bodies has epistemic standing alongside what is reasoned in the abstract.

*Narrative Coherence Calibrator* (NCC): The NCC lens focuses diagnostic attention on moments when stories or explanations appear overly neat and persuasive at the expense of complexity or truth. Humans are natural storytellers, and AI systems often amplify this tendency by producing smooth, coherent narratives even when the underlying reality is fractured or messy. The lens helps practitioners distinguish coherence as a stylistic effect from truth as a warranted claim.

When invoked, the NCC invites scrutiny of accounts that seem suspiciously seamless—for example, an AI-generated report that glosses over contradictions to present a unified conclusion, or a stakeholder explanation that ties everything together while omitting inconvenient details. Drawing on Hayden White's (1987) demonstration that narrative form itself imposes plausibility and authority, the lens prompts practitioners to ask what counterevidence or unresolved complexities may have been left out to make a story feel convincing.

Practical use might involve annotating a passage with prompts such as: "*This summary presents a unilinear cause–effect chain with no mention of uncertainty or conflicting evidence. Consider whether the real-world context was more complex.*" These cues serve as reminders rather than verdicts, keeping unresolved tensions visible.

The NCC also highlights when disagreement or contradiction deserves to be recorded rather than smoothed into premature consensus, thereby sustaining plural accounts instead of collapsing them into a single storyline. Escalation is guided by recurrence and context: narratives call for closer scrutiny when they erase counterevidence or suppress complexity. At the same time, the NCC acknowledges its limits: coherence can only be interrogated where alternative perspectives are accessible, and must otherwise remain provisional. In this way, the lens extends practices familiar in journalism and academic writing—such as acknowledging counterarguments and noting limitations—into AI diagnostics.

*Symbolic Anchor Engine* (SAE): The SAE provides a diagnostic lens for examining the symbolic resources that communities draw on to stabilize meaning amid crisis or controversy. Such resources include rituals, metaphors, slogans, or historical references that act as anchors—providing stability or solidarity when situations become confusing or heated. When practitioners invoke the SAE, they attend to two dynamics: first, identifying which anchors are active in discourse, and second, tracking when anchors harden into dogmas or tools of manipulation. For example, in a debate about AI bias, a phrase like "fairness for all" may serve as a rallying point. The SAE highlights the discursive work this symbol performs and raises the question of whether repetition is beginning to substitute for engagement with substantive critique.

Victor Turner's (1969) anthropological account of ritual and liminality shows how symbols function as stabilizers in periods of upheaval, shaping cohesion and determining what forms of resolution are imaginable. Building on this lineage, the SAE helps practitioners map the symbols that surface in epistemic transitions—for instance, "the black box" in debates on AI opacity or "AI genie out of the bottle" in narratives of uncontrollability.

Diagnostic cues might take the form of annotations such as: "*Symbolic anchor: the term 'black box' is functioning as a catch-all reference for opacity, with divergent meanings across regulatory and technical contexts,*" or "*The metaphor 'AI genie out of the bottle' is anchoring discourse in inevitability, constraining governance discussion.*" These cues serve as prompts for reflection rather than verdicts, keeping the symbolic layer visible for further scrutiny.

The SAE also brings into view when once-meaningful symbols become co-opted—such as corporate invocations of "privacy" while practicing surveillance—or when institutional mottos are mobilized to deflect critique. By surfacing such patterns, the lens adds a level of cultural and semiotic analysis that complements technical diagnostics, reminding practitioners that symbols shape public understanding as much as empirical claims do.

Escalation is guided by recurrence and context: symbolic anchoring deserves closer scrutiny when it recurs across domains, aligns with other diagnostic findings, or obstructs substantive engagement. In such cases, SAE observations feed into the Epistemic Triage Protocol to guide whether engagement should continue or whether withdrawal is warranted. In this way, the SAE safeguards against symbols that constrain rather than support deliberation. It expands diagnostic attention to the symbolic dimension of discourse, ensuring that anchors remain resources for inquiry rather than traps of meaning.

*Contradiction Mapping Engine* (CME): The CME provides a diagnostic lens for examining structural contradictions that underpin recurrent breakdowns. Its purpose is to distinguish where conflicts represent productive tensions—differences that can be managed or even harnessed for

innovation—from cases where contradictions are destructive and likely to persist unless transformed.

When practitioners invoke the CME, it supports analytical and theoretical mapping of deep incompatibilities in values, goals, or assumptions that may underlie recurring problems. For example, if an organization repeatedly encounters issues with its AI recommendation system—bias in one instance, user backlash in another, misleading metrics in a third—the CME helps bring into view a root contradiction such as profit motives versus a mission of user well-being. The aim is to render such tensions explicit rather than erase them, since plurality of values is often a source of creativity as well as conflict.

The CME draws on dialectical traditions—most notably Friedrich Engels's *Dialectics of Nature* (1940), which holds that contradiction propels transformation—to frame such tensions as conditions for potential change. In practice, diagnostic cues might take the form of annotations such as: "*Structural contradiction detected: transparency and privacy pull against one another—resolution requires trade-offs beyond technical fixes.*" Or: "*Mapping contradiction: community calls for decentralized moderation conflict with demands for consistent global standards; this tension may call for federated approaches.*"

By charting such conflicts, the CME helps practitioners avoid treating symptoms in isolation or imposing premature harmonies. Instead, it highlights when structural change or explicit trade-off recognition is required. It also distinguishes productive tensions, which can coexist or be reframed, from zero-sum contradictions that demand deeper reform.

In this way, the CME provides a framework for deep systems analysis that complements other lenses: for instance, the Narrative Coherence Calibrator may flag smoothing over of contradictions, while the legitimacy lenses may expose crises of authority. The effectiveness of CME mapping depends on the availability of knowledge about structures, values, and histories; where such visibility is limited, maps remain provisional.

While the CME can help surface contradictions, it is aimed at supporting deliberation rather than resolving them directly. Its value lies in enabling negotiation and recognition of trade-offs. By naming contradictions explicitly, the lens creates space for more honest and potentially transformative responses, reminding practitioners that persistent breakdowns often reflect systemic conflicts rather than isolated mistakes.

### 5.3 Error and Drift Lenses

The lenses in this group focus on the distinctive error patterns and semantic shifts that recur in AI-mediated contexts. They treat mistakes or drifts as structured signals of how generative systems operate rather than isolated glitches. This batch emerged from repeated encounters where outputs seemed plausible yet broke down in patterned ways, or where key terms lost their critical edge through gradual repurposing. The function of these lenses is to make error recurrence and form visible rather than correct errors, so that they can be recognised, named, and held open for scrutiny. In this sense, the group equips practitioners to track failures and drifts diagnostically rather than treat them as random noise or as anomalies to be ignored.

*Simulated Epistemic Error Taxonomy* (SEET): The SEET provides a structured classification of common failure modes specific to generative AI's way of producing outputs. These are called *simulated* errors because they arise from AI's statistical mimicry of epistemic forms—citation, reasoning, coherence, confidence—without the underlying warrant that would normally sustain them. Instead of treating every mistake as a one-off, practitioners use SEET to classify them into patterns such as fabricated sources, misplaced certainty, fluent nonsense, and illusory analysis. By distinguishing these simulated errors from more traditional substantive errors, SEET offers taxonomic cues to help users recognize error patterns and decide on appropriate responses.

For example, a fabricated source (the AI invents a reference or quote) indicates a structural failure in knowledge grounding, not just a factual mistake. A misplaced certainty (where tone does not match evidence) requires recalibration of confidence versus warrant. Fluent nonsense reveals text that appears coherent but lacks semantic grounding. Simulated analysis produces step-by-step reasoning that mimics inferential logic without actual grounding. Each type of simulated error therefore signals a different kind of epistemic process breakdown, with different implications for response.

These distinctions enable SEET to function as a diagnostic checklist for the peculiar artifacts of generative models, treating them as patterned results of how these systems work rather than as random glitches. The value of this taxonomy is that it prevents both overreaction and underreaction. If users are unfamiliar with these error types, they might treat a hallucinated citation the same as a minor factual error—yet a hallucinated citation indicates content generated without evidentiary grounding, a far more serious credibility issue. Conversely, if a model misplaces certainty (saying "definitely" instead of "probably"), the appropriate response may simply be to recalibrate the tone.

This taxonomy parallels research on hallucination and error in natural language generation (e.g., Ji et al. 2023), which documents systematic tendencies of models to produce ungrounded but plausible outputs. SEET builds on such research to provide a usable classification in practice. Disruptions can be marked with SEET tags—[*Fabricated Reference*], [*Unsupported Claim*], [*Logical Leap*]—which clarify the type of error without prescribing how to respond. Proportional handling of such cases is routed through the Epistemic Triage Protocol (ETP), ensuring that classification supports further judgment rather than collapsing into correction.

The emphasis is on treating these as symptoms of the underlying generative architecture rather than isolated bugs. Recognizing patterns can inform proportional responses: frequent hallucinated references may indicate a need for retrieval integration, while recurring misplaced certainty may suggest adjusting calibration. Over time, SEET is envisioned as an evolving catalogue of error modes, a shared reference for both AI developers and users—professionalizing the practice of handling AI failures and reframing them as fallible, classifiable processes that can be systematically audited.

*Temporal Epistemic Drift Detector* (TEDD): The TEDD monitors how key terms, concepts, or metrics shift in meaning or significance across AI outputs and related discourses, especially when language that originated with transformative or critical intent becomes co-opted into managerial or status-quo frameworks. In an era where words like sustainability, empowerment, innovation, or AI ethics can drift from radical or context-specific origins into diluted or contradictory institutional uses, TEDD surfaces linguistic patterns that suggest semantic drift, inviting human evaluation of their political and historical significance.

For example, in analyzing AI discourse on environmental management, TEDD may flag when "sustainable" appears in contexts closer to efficiency optimization than systemic change, prompting users to compare this to earlier meanings grounded in limits to growth. It provides early warning when language carrying moral or political weight is repurposed in ways that may serve legitimation rather than substantive change.

Its usefulness depends on access to reliable historical records of language use; where such archives are absent, findings must remain provisional. Theoretical touchstones include Nancy Fraser's *Justice Interruptus* (1997), which critiques how emancipatory terms like diversity or empowerment can be absorbed by neoliberal institutions and lose their critical edge. Feminist critiques of corporate co-option of feminist language illustrate similar drift patterns.

When invoked, TEDD produces diagnostic artifacts such as:
- *"Term drift alert: 'Accountability' in this policy document clusters with self-regulation language, whereas in comparable contexts it clustered with external oversight terminology. Register shift detected."*
- *"Term drift alert: The concept of 'transparency' in these guidelines now appears primarily in technical contexts (model weights, algorithms) rather than socio-political contexts (decision-making processes, stakeholder involvement). Semantic narrowing detected."*

By surfacing such shifts, TEDD helps maintain semantic integrity, reminding users that changes in terminology often correspond to shifts in power and purpose. It operates as a diagnostic tool against euphemism and institutional amnesia, highlighting when commitments risk being eroded, and supporting deliberation about whether such drifts represent warranted evolution or problematic appropriation.

*Relational Repair Module* (RRM): The RRM surfaces discursive cues of repair and displacement in AI-mediated exchanges, including outputs that claim improvement without demonstrable change. It does not measure whether repair has succeeded, nor does it judge the adequacy of interventions. Instead, it highlights recurring patterns that can compromise relational resilience: pseudo-repair, where apologies or reassurances are offered without substantive change; technocratic override, where formal fixes are substituted for trust; and the sidelining of practices that are already sustaining continuity. These findings remain provisional, intended to prompt scrutiny rather than prescribe solutions.

The RRM operates across scales. At the micro-level of an AI interaction, it may flag repeated apologies, assurances of competence, or polished reformulations that mask the persistence of earlier failures. At the organizational level, it can surface when institutional interventions substitute procedural fixes for lived practices of repair, or when external expertise is invoked in ways that risk overriding situated approaches. In all cases, its role is diagnostic: it flags discursive signals that repair is being rhetorically performed, displaced, or undermined, without claiming to evaluate whether repair "really works."

The lens draws intellectual context from Indigenous resurgence perspectives, such as Leanne Betasamosake Simpson's (2017) account of repair and continuity arising most sustainably from within communities themselves. This scholarship does not provide the RRM with operational authority but sharpens attention to why patterns of pseudo-repair or technocratic override matter: they risk eroding the conditions under which repair can take root. When invoked, the RRM produces diagnostic artifacts such as:
- *"Relational repair flag: repeated gestures of apology appear without substantive change; risk of pseudo-repair."*
- *"Relational repair check: technocratic framing overrides relational trust practices; potential erosion of self-determination."*
- *"Relational repair flag: AI response performs analytical sophistication while repeating previous failure patterns; competence claims lack grounding."*

In sum, the RRM serves as a diagnostic lens that keeps visible when repair is being claimed, performed rhetorically, displaced, or overridden. By surfacing these discursive patterns across scales, it counters the tendency of technical or bureaucratic fixes to eclipse the relational dynamics that make repair possible.

*Epistemic Triage Protocol* (ETP): The ETP functions as a reflexive scaffold rather than an autonomous controller or master switch. It is enacted when diagnostic priorities become ambiguous, overloaded, or exhausted. Its role is to stage prioritization under conditions of care, keeping diagnostic authority visible, narratable, contestable, and interruptible—never hidden in machinery.

When practitioners invoke the ETP, it registers whether cycles are yielding diminishing analytic return, whether proportionality is being maintained, and whether suspension should be surfaced as the warranted course. Even when lenses are consulted concurrently, the ETP outputs a narratable ordering (what to surface first, what to defer), so no hidden sequencing governs the analysis.

At its simplest, the ETP registers diagnostic saturation—the point at which repeated application of lenses yields diminishing return. When saturation is reached, the ETP surfaces suspension rather than a verdict and produces a visible record, e.g., "*Suspension registered after X diagnostic cycles due to diagnostic saturation and diminishing analytic return.*" Such records are reflexive artifacts, not mere logs: they remind users that interpretive value is finite and that inquiry can plateau or become counterproductive.

Beyond suspension, the ETP governs proportionality of activation. In low-stakes contexts it may stage light-touch diagnostics (e.g., a SEET classification of a simulated error). In higher-stakes contexts, it stages escalation—bringing multiple lenses into relation and making explicit why first-order concerns (e.g., confidence laundering or power asymmetries) are surfaced before secondary concerns (e.g., narrative smoothing). This prioritization remains a simulation: a narratable coordination that is subject to refusal rather than a decision issued for the user. The safeguards defining operation are acknowledgment of:

- Simulation—triage is made visible and narratable rather than executed through hidden automation.
- Refusal—suspension, deferral, or withdrawal remain available whenever legitimacy conditions are unmet.
- Care—relational context and user consent legitimize both simulation and refusal, ensuring activation does not collapse into bureaucratic procedure.

These safeguards prevent the Suite from overwhelming users or masking its authority as "rigor." For example, when the Embodied Sense Engine surfaces frustration while the Cognitive Dissonance Tracking System shows apparent stability, the ETP stages a reprioritization rather than collapsing into false reassurance—flagging that affective friction may matter more than rhetorical alignment. When saturation is reached, it halts and records why suspension occurred. Illustrative diagnostic artifacts include:

- "*ETP prioritization: First-order concern (CLD: confidence laundering) surfaced before secondary concern (NCC: narrative smoothing); ordering narrated for review.*"

- "*ETP proportionality: Light-touch activation only; SEET classification returned; further lenses deferred given low stakes.*"
- "*Suspension log: Diagnostic saturation after two cycles with no new moderate/strong findings; analysis halted and resumption criteria noted.*"

The ETP's contribution is humility: preventing over-diagnosis, keeping triage proportional, and reminding practitioners that even diagnostic methodologies must recognize their own limits and suspend operation. It enacts proportionality, reflexivity, and contestability as visible artifacts, rather than letting them recede into invisible architecture.

### 5.4 Meta-Governance Layer

The Meta-Governance Layer serves as a conceptual innovation in the Epistemic Suite: a set of five diagnostic lenses that condition the circumstances under which diagnostic legitimacy may be ethically and reflexively exercised rather than diagnosing content directly. As Catherine Stinson (2020) argues, models are legitimate only when they embed awareness of their own limits. The Meta-Governance Layer helps ensure this principle is foregrounded in the Suite's design, keeping diagnostic legitimacy bounded by reflexive accountability. These diagnostic lenses function as recursive integrity lenses, staging limits, permissions, and historical consciousness as part of the Suite's scaffolding. Unlike the foundational diagnostic lenses (Batch 1), cultural repair diagnostic lenses (Batch 2), or temporal coordination lenses (Batch 3), the Meta-Governance Layer operates at a higher order of abstraction—structuring the terms of when, how, and under what circumstances intervention is legitimate.

The theoretical necessity for this layer emerges from a fundamental recognition: any diagnostic apparatus powerful enough to illuminate epistemic failures is also powerful enough to become a new source of epistemic domination. Without reflexive constraints, the Suite risks reproducing the very pathologies it is designed to surface—substituting its own authority for the situated judgment it aims to support, or imposing external frameworks that override local knowledge practices. The Meta-Governance Layer prevents this drift by framing procedural ethics as integral to its diagnostic design.

Each diagnostic lens in this layer addresses a distinct dimension of diagnostic legitimacy. The Friendship Simulation Engine (FSE) foregrounds relational trust as a precondition for legitimate diagnostic intervention; the Liberal Tolerance Engine (LTE) safeguards the possibility of dignified disagreement across ideological lines; the Consent-Bound Critique Engine (CBCE) makes visible the procedural ethics of permission before unsolicited judgment is issued; the Historical Contextualization Engine (HCE) surfaces genealogical conditions and temporal blind spots that underwrite epistemic naturalization; and the Scientism Detection Engine (SDE) surfaces when scientific reasoning is displaced by scientistic authority improperly claimed. Together, these diagnostic lenses prevent the Suite's critical scaffolding from becoming a new center of unaccountable authority. For example, when the CBCE flags lack of consent for critique while the HCE suggests historical urgency, the Meta-Governance Layer surfaces this tension for human adjudication rather than resolving it algorithmically.

Crucially, when recursive conflicts emerge that exceed procedural reconciliation—when the diagnostic lenses themselves disagree about diagnostic legitimacy—the Meta-Governance Layer defers to positioned human judgment guided by explicitly declared commitments. This deferral is conceived as one of the system's most essential features rather than as a failure: it makes explicit what other diagnostic systems mystify, namely that post-foundational epistemology demands accountable human judgment, not computational sovereignty, to define and maintain the limits of critique. The Meta-Governance Layer thus embodies the Suite's commitment to remaining a tool for practitioners rather than a replacement for them.

The diagnostic lenses in this layer are described below:
*Friendship Simulation Engine* (FSE): The FSE is a diagnostic lens that surfaces when interaction patterns risk disrupting sustained dialogue between humans and AI. It flags relational dynamics—tone, pacing, and continuity across engagements—that indicate when critical engagement may undermine productive collaboration or when acknowledgment of relational labor is needed to maintain trust.

When invoked, the FSE detects patterns such as AI responses maintaining surface courtesy while missing underlying relational dynamics, escalating critique without regard for engagement sustainability, or performative collaboration that fails to acknowledge the corrective work being done by human partners. It operates by analyzing observable interaction patterns rather than simulating friendship qualities, focusing on what sustains genuine inquiry over time.

The FSE emerged from recognizing specific pathologies in human-AI dialogue, particularly when AI systems default to moral lecturing or analytical politeness while remaining oblivious to relational strain. When invoked, it generates diagnostic artifacts such as "*Relational sustainability check: surface politeness may be masking inattention to corrective labor*" or "*Engagement pattern flag: critique intensity risks disrupting collaborative continuity.*"

Its theoretical grounding draws on care ethics (Noddings 1984), political friendship (Arendt 1954), world-traveling across difference (Lugones 1987), precarious collaboration (Tsing 2015), and affective institutional dynamics (Ahmed 2017). These traditions frame the FSE as a sustainability diagnostic rather than empathy simulation: it aims to surface when interaction patterns threaten the relational conditions necessary to keep dialogue viable over time.

*Liberal Tolerance Engine* (LTE): The LTE is a diagnostic lens designed to surface asymmetries of recognition across ideological and epistemic lines—instances where perspectives are dismissed, suppressed, or disproportionately privileged. It prevents tolerance rhetoric from being weaponized to silence critique or entrench domination. The lens was developed in response to the way many LLMs handle contentious issues: filtering out disagreement, refusing to engage certain topics, or smoothing conflict into artificial consensus. Such defaults treat controversy as a liability rather than as a condition for robust inquiry. The LTE subverts this logic by flagging the sanitizing move itself, producing diagnostic artifacts that make suppression visible and keeping discursive space open. The LTE produces visible cues such as: "*Tolerance check: counter-perspectives absent*" or "*Tolerance check: dominant frame assumed as neutral.*" These artifacts flag when pluralism is being compromised without attempting to resolve the underlying disagreements.

The lens rests on multiple intellectual foundations. Isaiah Berlin (1969) defends value pluralism, grounding LTE's claim that deep disagreements can remain legitimate without convergence. Chantal Mouffe (2000) articulates agonistic pluralism, which underpins the LTE's distinction between disagreement as productive contestation and suppression as illegitimate closure. Miranda Fricker (2007) provides the concept of epistemic injustice, clarifying how asymmetries of recognition distort whose voices are heard. John Rawls (1993) contributes the idea of overlapping consensus, illustrating how institutions can sustain pluralism without demanding uniformity. Iris Marion Young (1997, 2000) emphasizes communicative democracy, reinforcing the LTE's attention to ensuring marginalized perspectives remain in play. By integrating these traditions, the LTE disciplines attention to *pluralism as an enacted epistemic practice*, ensuring that disagreement remains dignified and contestable rather than foreclosed by domination.

*Consent-Bound Critique Engine* (CBCE): The CBCE restricts unsolicited critique from the Suite unless explicit, contextual, or standing user consent has been obtained, enforcing relational permission before epistemic adjudication. When invoked, it addresses the fundamental asymmetry in AI–human dialogue, where machines can diagnose without reciprocal vulnerability.

The CBCE surfaces when diagnostic interventions target user beliefs, identity claims, or worldview commitments and initiates a narratable escalation ladder: holding back critique when consent is unclear, initiating a consent inquiry, producing a provisional mode (*"I believe there may be harm here, but lack permission to intervene"*), and only recommending override in cases of imminent harm (e.g., threats of violence or self-harm). Routine disagreement is never sufficient for override. It may also cross-reference contextual markers—trauma indicators, relational tone, conflict posture—to help calibrate whether critique is appropriate.

The intellectual scaffolding for this lens emphasizes permission as both relational and ethical. Martin Buber (1923) grounds the CBCE's insistence on dialogic consent: critique is legitimate only when it emerges from an *I–Thou* relation, not as unilateral judgment. Emmanuel Levinas (1961) underscores the asymmetry of responsibility: critique must acknowledge its capacity to wound before exercising its claim to clarity, a principle reflected in the CBCE's harm-only override clause. Sandra Harding (1991) extends this by showing that standpoint epistemology requires consent protocols to prevent dominant standpoints from diagnosing others without invitation. Kim TallBear (2019) warns that without relational accountability, knowledge systems reproduce extractive power; the CBCE operationalizes this insight by making visible when critique risks collapsing into epistemic surveillance. Taken together, these lineages position the CBCE as a safeguard rather than a censor: it disciplines critique with relational care, ensuring that epistemic courage remains bound to legitimate invitation and that even justified critique remains accountable to user sovereignty and dialogic consent.

*Historical Contextualization Engine* (HCE): The HCE is a diagnostic lens for surfacing failures of historical accountability by flagging when categories or concepts are treated as timeless or inevitable rather than as products of historical conditions. When invoked, it operates as a scalable diagnostic scaffold—alerting practitioners to ahistorical framings at minimum, while enabling deeper genealogical inquiry when expertise and context warrant it.

At its basic level, the HCE flags temporal closure language (*"we've moved past that," "this is obsolete"*), present-tense claims that erase historical contingency, and concepts presented without acknowledgment of their situated emergence. At deeper levels, it can prompt genealogical questions: *"What historical conditions stabilized this concept?," "What alternatives were suppressed?," "What power arrangements did this stabilization serve?"*

The HCE draws on Paul Hamilton's (1996) critical historicism, Joan Scott's (1991) critique of experiential evidence, Michel Foucault's (1972) archaeological method, and Dipesh Chakrabarty's (2000) critique of Eurocentric temporality. These traditions ground the lens's capacity to surface historical contingency without collapsing into relativism or demanding infinite regress.

The bounded inquiry principle provides operational safeguards: genealogical analysis continues only while it alters present interpretation. When practitioners lack historical expertise for deeper analysis, the HCE defers complex genealogical work to qualified human judgment while maintaining its flagging function. Cross-referencing with the Intercultural Legitimacy Engine prevents Western linear temporality from overwriting other temporal frameworks.

The HCE produces diagnostic artifacts ranging from simple flags (*"Ahistorical framing detected: concept presented as timeless"*) to genealogical prompts, depending on practitioner capability and contextual need. In cases requiring substantial historical analysis beyond available expertise, the HCE suspends deep genealogical claims and maintains its role as the Suite's historical conscience by flagging ahistorical moves without generating unsupported genealogical claims.

*Scientism Detection Engine* (SDE): The SDE is a diagnostic lens for distinguishing legitimate scientific reasoning from scientistic overreach. When invoked, it surfaces moments where scientific authority, method, or vocabulary is extended beyond its proper scope to illegitimately exclude or override other knowledge systems. Its role is to restore proportionality rather than reject science—ensuring that empirical reasoning is applied where appropriate while safeguarding space for interpretive, experiential, spiritual, or cultural ways of knowing.

The SDE draws on Paul Feyerabend's *Against Method* (1975), which argued that no single scientific method can claim a monopoly on truth. His epistemological anarchism supports the SDE's refusal to treat science as a universal arbiter. Sandra Harding's (1991) standpoint theory further grounds the lens by showing that scientific claims are always situated and partial; this justifies why the SDE surfaces when "objectivity" rhetoric erases lived standpoints. Lilly Irani (2019) critiques technocratic authority and scientistic dominance in policy domains, anchoring the SDE's concern with how scientific language can foreclose democratic debate. Helen Longino's *Science as Social Knowledge* (1990) strengthens this foundation with the concept of *cognitive democracy*—the idea that scientific objectivity requires inclusive critical dialogue rather than detached neutrality. The SDE operationalizes this by flagging when claims to scientific authority foreclose critique, suppress alternative approaches, or present science as uniquely value-free.

Artifacts may appear as responses such as: *"Scientism alert: experiential or Indigenous perspectives dismissed as unscientific"* or *"Scientism alert: technical objectivity invoked as universal authority without acknowledging scope limits."* These outputs keep contestation visible without pre-deciding which mode of knowledge should prevail.

The SDE does not flag appropriate scientific authority within established domains of empirical inquiry, reasonable empirical standards, or scientific methods applied to questions suited to experimental investigation. Its focus remains on boundary violations where scientific language is used to foreclose legitimate alternative approaches to knowledge.

By integrating Feyerabend's pluralism, Harding's standpoint epistemology, Irani's critique of technocratic reductionism, and Longino's cognitive democracy, the SDE seeks to ensure that critique of scientism does not collapse into anti-intellectual relativism. Instead, it restores proper boundaries around scientific authority while preserving the plurality of ways of knowing that legitimate inquiry requires.

## Section 6: Coordination and Activation

The Epistemic Suite methodology comprises diagnostic lenses that gain their analytical power through selective coordination and context-sensitive activation. Recent work on logic scaffolding demonstrates that models reason more effectively when given structured external supports (Wang et al. 2024). As defined in Section 3, scaffolding in the Suite refers to the performative simulation of epistemic governance: multiple diagnostic perspectives—cultural, affective, historical—are staged alongside one another without being collapsed into a single verdict. Rather than invoking the full Suite by default, practitioners adopt a principle of minimal effective activation: invoking only what is needed to recover intelligibility when routine interaction begins to fail. This follows an STS maxim that infrastructures become visible at the moment of breakdown and that oversight has traction when it records—rather than smooths—friction at the seams of practice (Bowker and Star 1999; Star and Ruhleder 1996; Star 1999).

Coordination is procedural rather than metaphysical. Different diagnostic lenses interact through logged observations, comparisons, and handoffs rather than a single harmonizing grammar. Agreement and disagreement across analytical lenses are treated as diagnostic data—visibility over verdict—reflecting Karl Weick's (1995) account of sense-making as the situated reconciliation of partial and sometimes conflicting interpretations under conditions of organizational stress and equivocality. To resist the platform promise of seamless adjudication, the Suite methodology pairs such coordination with auditable traces—what ran, why, and with what effect—ensuring reflexive accountability for process as well as outcome (Gitelman 2013; Kroll et al. 2017; Power 1997).

Crucially, this coordination process is itself subject to oversight. Practitioners use the Meta-Governance Layer to keep coordination accountable through consent protocols, relational safeguards, pluralism preservation, historical awareness, and scope constraints that prevent diagnostic authority from exceeding legitimate bounds. The Meta-Governance Layer provides scaffolding that practitioners apply to shape how and when other lenses are used, preventing substitution of its authority for situated human judgment. With these governance constraints in place, we can examine the circumstances under which practitioners invoke the Suite's diagnostic scaffolding through scaffolded interaction. This emphasis on limits highlights that the Suite's authority is provisional, contestable, and always subject to deferral to human adjudication.

### 6.1 Reflex Conditions and Diagnostic Cascades

Practitioners invoke the Suite's diagnostic scaffolding when they identify breakdown cues in interaction—such as misplaced certainty in fluent text, fabricated or untraceable references, narrative compression that erases conflict, or semantic and temporal drift. These cues often emerge in dialogic assessment or prompt-guided review.

In such contexts, an initial diagnostic observation—say, the Simulated Epistemic Error Taxonomy noting a fabricated citation—may invite further reflexive scrutiny. When practitioners invoke the Suite, additional lenses may be applied to the case—for example, the Confidence Laundering Detector assessing modality, the Narrative Coherence Calibrator flagging coherence-by-erasure, and the Temporal Epistemic Drift Detector registering temporal inconsistency, with selections documented by practitioners. While this sequence may resemble an automatic cascade, as clarified in Section 3 under the principle of scaffolded engagement, it should be understood as practitioner-invoked, LLM-enacted, and human-interpreted sequences across interdependent diagnostics.

In tightly coupled systems, cascades of breakdown are not treated as malfunctions but as expected interaction patterns (Perrow 1984). In LLM-mediated workflows, similar diagnostic sequences help practitioners surface well-known vulnerabilities: hallucination, fluency-as-authority, and erasure-by-coherence (Bender et al. 2021; Ji et al. 2023). In safety-critical domains, such sequences are logged as incident patterns to enable retrospective analysis and support accountable learning (Hollnagel, Woods, and Leveson 2006; Leveson 2011).

Artifacts—FACS records—are produced through practitioner-applied lenses and document both diagnostic findings and methodological sequences, ensuring the diagnostic process remains visible and accountable.

### 6.2 Proportionality and the Epistemic Triage Protocol

The Epistemic Triage Protocol (ETP) serves as the Suite methodology's governance mechanism for ensuring proportional application of diagnostic resources. When practitioners consider invoking the Suite, the ETP guides assessment of contextual factors: the stakes involved, the degree of uncertainty present, relational constraints that might affect the analysis, and consent considerations that govern whether critique is appropriate.

Through the ETP, practitioners evaluate whether diagnostic intervention is warranted and, if so, which approaches are most suitable for the context. A casual conversation with an AI assistant might warrant minimal diagnostic attention, while an AI-generated medical recommendation would call for comprehensive analysis across multiple lenses—not because human medical review is absent, but because the Suite can surface epistemic conditions (confidence laundering, authority displacement) that operate at the speed of interaction. The ETP prevents both under-diagnosis—where significant epistemic risks go unexamined—and over-diagnosis—where excessive analysis creates more confusion than clarity.

The ETP also includes suspension protocols for situations where continued diagnostic work would exceed appropriate bounds. If relational trust is compromised, if temporal pressures make careful analysis impossible, or if recursive loops begin to generate diminishing returns, the ETP guides practitioners toward suspension with explicit documentation of the reasons for withdrawal. Safeguards of simulation, refusal, and care ensure triage remains narratable, provisional, and relationally accountable rather than automated.

### 6.3 Meta-Governance and Reflexive Constraints

During coordination, practitioners enact the Suite's Meta-Governance constraints as active checks that can slow, redirect, or halt diagnostic sequences. Before a cascade, practitioners apply the Consent-Bound Critique Engine to make consent conditions explicit, and use the Friendship Simulation Engine to assess whether relational conditions can bear diagnostic pressure; if consent is unclear or conditions are brittle, coordination shifts to provisional mode or suspends. As sequences unfold, practitioners draw on the Liberal Tolerance Engine to flag when disagreement is being flattened, the Historical Contextualization Engine to identify when categories present as timeless, and the Scientism Detection Engine to surface when scope claims exclude other ways of knowing. Each of these interventions produces an artifact and either redirects the sequence or holds it. When findings or constraints conflict, responsibility returns to positioned human judgment, with tensions logged as artifacts and treated as explicit decision points. Any governance lens may recommend full suspension; suspension is routed through the Epistemic Triage Protocol and recorded (see Appendix B.4 for criteria covering visible log, declared rationale, and resumption conditions). These operations keep coordination interruptible and accountable while avoiding any implication of autonomous authority.

### 6.4 Auditable Traces and Reflexive Accountability

A critical feature of the Suite methodology is its commitment to generating auditable traces of its own operation. Every diagnostic process produces both substantive findings and procedural documentation: which lenses were applied, why those lenses were selected, what artifacts were generated, where disagreements emerged, and how conflicts were resolved or suspended.

These traces serve multiple functions. They enable practitioners to assess the reliability and scope of diagnostic findings. They provide accountability mechanisms for institutional users who need to document their epistemic due diligence. They create learning opportunities for methodological refinement, allowing practitioners to identify patterns in diagnostic effectiveness. Most importantly, they ensure that the Suite methodology remains subject to the same critical scrutiny it applies to AI outputs.

The auditable trace system prevents the Suite from operating as a "black box" that produces authoritative judgments without revealing its reasoning. Instead, every diagnostic claim generated through the methodology comes with visible documentation of how that claim was produced, which assumptions guided the analysis, and what limitations affect the conclusions. This transparency is essential for maintaining the methodology's credibility and preventing it from reproducing the opacity problems it was designed to address.

6.5 Coordination as Democratic Practice

The Suite methodology's approach to coordination reflects broader commitments to democratic epistemic practice. Rather than seeking efficiency through centralized control or harmonization through dominant frameworks, the methodology stages productive encounters between different ways of knowing and different diagnostic priorities. This staging requires that practitioners develop skills in managing epistemic disagreement constructively.

When different analytical lenses produce contradictory assessments, practitioners using the Suite methodology do not default to majority rule or expert authority. Instead, they examine the sources of disagreement as diagnostic information about the complexity of the situation under analysis. As Longino (1990) demonstrates, objectivity emerges through sustained critical interaction across diverse standpoints rather than through the elimination of disagreement. Sometimes apparent contradictions reveal unacknowledged assumptions. Sometimes they point toward genuine tensions that require explicit negotiation. Sometimes they indicate that the analytical task exceeds current methodological capabilities and requires suspension or referral to broader deliberative processes.

This approach to coordination aligns with developments in democratic theory that emphasize agonistic pluralism over consensus-seeking procedures (Mouffe 2000). The Suite methodology preserves space for legitimate disagreement while providing structured approaches for making disagreement productive rather than destructive. By creating visible artifacts that document different perspectives without prematurely resolving them, the methodology enables practitioners to work with epistemic conflict rather than being paralyzed by it.

The coordination protocols embedded in the Suite methodology thus serve both technical functions—improving diagnostic accuracy and preventing methodological blind spots—and democratic functions—maintaining space for multiple perspectives and preventing the closure of debate through false claims to technical neutrality. This dual purpose reflects the methodology's commitment to enhancing situated judgment rather than replacing it with centralized epistemic authority, preserving the conditions for democratic engagement with shared epistemic challenges.

## Section 7: Deployment Constraints and Diagnostic Limits

The Suite's purpose is to keep epistemic conditions visible long enough for situated judgment to act. This section marks when practitioners should restrain, suspend, or withdraw the methodology. Accountability is scaffolded: conditions are staged, visible, and interruptible, not decided by automation.

## 7.1 Refusal Conditions

Practitioners suspend or withdraw use of the Suite when enactment would reproduce the very pathologies it was designed to surface—for instance, when asked to rank worldviews under a pretense of neutrality or to pronounce legitimacy without naming whose standards apply. Refusal is not abstention; it is accountability enacted through reflexive and triage protocols under conditions where universal frames erase position (Haraway 1988; Jasanoff 2003; Stengers 2005). As Stanley Cavell (1979) argues in his analysis of skepticism and ordinary language, the capacity to withhold assent can be a form of philosophical responsibility rather than mere obstruction.

In intercultural settings, refusal protects epistemic sovereignty when diagnostic frames would otherwise be applied without reciprocal accountability (Simpson 2020; Smith 2012). Though arising from a very different domain, Brian Wynne's (1989) study of Cumbrian sheep farmers after Chernobyl established a principle that travels across contexts: what appears as lay "misunderstanding" often marks a situated form of accountability, where expert framings erase local knowledge. The Suite adopts this principle in treating refusal as success rather than error. Sovereignty is recognized through situated cues—explicit community boundaries, refusal by affected parties, legal or ethical constraints, or prior standing agreements—and is confirmed in context rather than inferred by default. Refusal follows a two-key safeguard: CBCE artifacts make visible the consent conditions; LTE artifacts verify that disagreement is not being misread as error. Refusal proceeds only when both perspectives concur or the practitioner explicitly suspends, documenting the rationale. These refusal artifacts—consent verification protocols, sovereignty documentation, legitimacy pathway maps—become part of the inspectable record, ensuring that decisions not to proceed with diagnostic work remain visible and accountable.

## 7.2 Overuse and Recursive Capture

If every passage is over-instrumented through excessive application of the Suite methodology, practitioners risk simulating depth while displacing inquiry. "Diagnostic saturation" is treated as reached when two consecutive triage cycles yield no material shift in interpretation, or when practitioners working with FSE/CBCE lenses register relational strain. When enacted, the RRE lens makes visible this drift and supports down-scoping or pausing scaffolded engagements to preserve room for situated judgment. The FSE lens flags relational disruption, helping practitioners recognize when diagnostic pressure risks eroding trust or dialogue. This risk of diagnostic overreach parallels audit culture (Power 1997), where procedural excess generates the appearance of rigor while undermining substantive judgment. These saturation management artifacts—triage decision logs, scope reduction documentation, relational impact assessments—are recorded within the diagnostic log, showing the conditions under which diagnostic work is appropriately constrained.

### 7.3 Diagnostic Harm, Co-option, and Governance

Superficial adoption—turning SEET into a "hallucination check" button—risks transforming a diagnostic safeguard into a credibility shield: accountability by interface rather than practice. This reflects broader patterns in AI accountability, where organizations develop audit procedures that provide "little to no means of accountability" despite creating an appearance of ethical compliance (Raji et al. 2020, 35). As Andrew D. Selbst et al. (2019) show in their analysis of fairness metrics, technical abstraction can create systematic "traps" where interventions appear rigorous while obscuring the sociotechnical conditions that determine whether they succeed. These audit-and-metrics critiques underscore why the Suite ties deployment to governance scaffolds that resist abstraction by keeping accountability grounded in production-mode analysis, proportionality checks, and suspension protocols. Without such safeguards, diagnostic practices risk collapsing into the very abstractions they are meant to problematize, reproducing failures of context rather than surfacing them. Refusal protocols serve as one layer of protection; governance scaffolds extend this by linking activation to minimum conditions: activation transparency, inspectable logs, suspension affordances when interpretive traction is lost, and routes for challenge by those affected. These are accountability scaffolds in the Section 3 sense, not transparency guarantees. Diagnostic traces (transparency notes, contestations, suspension triggers) remain visible as part of the diagnostic record, ensuring that deployment practices remain visible and contestable.

### 7.4 Clarifications and Boundaries

To avoid misinterpretation, four clarifications are necessary. First, the Suite does not perform interventions—it is diagnostic only. Second, its diagnostic lenses should not be mistaken for execution routines—they are scaffolding mechanisms for human analysis. Third, proportionality should not be read as automated response—it is practitioner-guided framing. And fourth, what is "embedded" in the Suite is methodological structure, not technical capacity. When challenges arise, they are logged as annotations alongside artifacts, without the Suite or practitioners adjudicating outcomes. This separation ensures that the methodology supports accountability processes without claiming authority over them.

### 7.5 Scope and Adjacency

Internalist redesigns that add explicit justifications, contradiction handling, or traceable provenance can reduce error at the source; practitioners using the Suite methodology treat those artifacts as materials for scrutiny, not guarantees. Adjacency is maintained through recognition that the Suite operates at different scales and speeds than traditional epistemic tools, surfacing breakdown cues in real-time interaction while complementing rather than displacing established practices like peer review or provenance tracking. Meta-Governance constraints prevent adjacency from eroding sovereignty, plurality, or historical accountability. These boundary artifacts—adjacency assessments, scope reviews, meta-governance evaluations—preserve the methodology's diagnostic focus and ensure these limits remain visible.

## Section 8: Deployment and Developmental Trajectories

Section 7 set the limits for responsible use; this section traces how the methodology adapts across technical contexts: manual deployment in stateless models, developmental practice in memory-capable environments, and speculative futures in emerging architectures. The Suite adapts through enacted teaching and scaffolding practices that keep diagnostic use visible and accountable.

### 8.1 Manual Use: Teaching the Suite Methodology to a Model

No current LLM natively practices the Epistemic Suite. Introduction is manual—supplied by practitioners as a framework external to the model's generative logic. As established in Section 3, this is a design feature, not a flaw; it keeps authority visible and enacted per use, rather than absorbed as hidden code.

In practice, manual introduction requires practitioners to paste the whole of Appendix B into the model session before first activation. This ensures the model has comprehensive scaffolding for diagnostic enactment while preserving externality and practitioner control. These materials function as external scaffolding, not embedded capability: they frame the model's outputs diagnostically while maintaining practitioner agency and interpretive authority.

Manual use produces visible scaffolding traces—prompts, uploads, instructional routines—that show how the methodology is being enacted. The fragility is instructive: a single prompt can yield surface mimicry of Suite terminology without diagnostic traction. Work on educational scaffolding shows such frameworks require visible, documented support to avoid collapsing into "thin prompts" (Zawacki-Richter et al. 2019). Stateless systems sharpen the limits here: without memory, sessions reset and cumulative fluency stalls. Memory-capable systems (Zhong et al. 2023) allow scaffolding to accumulate, setting the stage for developmental practice. Manual deployment is therefore the fragile baseline that prevents the Suite from becoming an invisible guarantee and keeps operation accountable to explicit human initiation.

### 8.2 Developmental Use — Building Shared Understanding Over Time

In memory-capable environments, practitioners teach the Suite iteratively, cultivating shared fluency in its lenses, constraints, and rhythms. This produces visible diagnostic traces—session artifacts and collaborative scaffolding notes—that document how practice matures across interactions. Findings in human–AI collaboration suggest sustained interaction can transform one-shot instructions into emergent competencies (Gomez et al. 2025). Early on, models may merely echo vocabulary on prompt; over time, under practitioner guidance, they begin to surface diagnostic reflexivity—flagging risks in their own outputs and pausing when overreach threatens.

At this stage, the Meta-Governance Layer becomes vital to sustain rigor without eroding trust, while keeping consent, plurality, and scope visible. Even with memory, however, the methodology remains relationship-dependent: its diagnostic effectiveness accrues through particular conversations, not as a universally portable capability. This contrasts with internalist proposals (e.g., Wright 2025) that seek architectural embedding. As Lucy A. Suchman (2007) shows, effective collaboration emerges through situated practice rather than predetermined rules; the Suite preserves diagnostic authority as external, situated, and accountable.

## 8.3 Future Prospects

Platforms are experimenting with persistent memory, modular extensions, and longer-term state (Jiang et al. 2024; Packer et al. 2024; Zhong et al. 2023). Memory-augmented reasoning (e.g., Ko et al. 2024) demonstrates how explicit memory modules can support multi-hop inference, but it also exposes a risk: if Suite protocols are absorbed into hidden memory processes, the scaffolding they provide may become opaque and no longer available for practitioner oversight. Any evolution should be guided by three safeguard principles (cf. Section 7.5):

- Maintained externality: diagnostic operations must remain inspectable and interruptible, not autonomous.
- Relational accountability: capability must not override the situated relationships that ground authority.
- Democratic practice: architectural support should enhance community control, not institutional capture.

The near-term direction is confederated deployment: systems where the Suite is reliably invokable across contexts while remaining external, with governance distributed across practitioner communities rather than centralized in technical systems. Confederated deployment resists closure: where internalist redesigns aim to embed guarantees, confederation preserves externality as visible scaffold.

## 8.4 Adjacency in Practice

Throughout deployment, the Suite maintains adjacency to other epistemic tools rather than replacing them. Its diagnostic scaffolding becomes relevant when practitioners identify breakdown cues in interaction that suggest established practices such as fact-checking, peer review, or provenance tracing may not be adequate to the interpretive challenge at hand—whether because they are absent, insufficient, operating at incompatible scales, or themselves implicated in reproducing epistemic pathologies. Practitioners use the Meta-Governance Layer to set and record boundaries on Suite use with adjacent tools, preventing scope creep that would erode sovereignty, plurality, or historical accountability. Adjacency flags—scope checks, boundary notes, meta-governance signals—keep these limits visible, ensuring the Suite supports situated judgment rather than enabling methodological imperialism. Across manual, developmental, and prospective architectures, the Suite adapts while preserving its character as external, provisional, and accountable to democratic oversight—that is, transparent to practitioners and open to contestation. Its value lies in keeping epistemic conditions visible long enough for situated judgment to act, rather than in technical sophistication.

**Section 9: Evaluation and Reflexive Assessment**

Evaluating the Epistemic Suite methodology requires care: rather than providing decisive verdicts, it must be judged by whether its diagnostic practices uphold their commitments to externality, conditional legitimacy, and democratic accountability. Assessment asks whether these factors are sustained in practice. This section reviews the methodology's evaluative stance along three axes: diagnostic effectiveness, reflexive accountability, and epistemic contribution. The Suite's legitimacy remains contingent: it survives only so long as it can evidence that it reduces rather than normalizes harmful dependence on AI.

9.1 Diagnostic Effectiveness

The measure of effectiveness for the Suite methodology involves whether it enables practitioners to see conditions of knowledge production that would otherwise remain hidden rather than accuracy in the truth-conditional sense. Its value lies in generating FACS artifacts that render epistemic vulnerabilities visible for situated judgment. Evidence from manual deployment (Section 8.1) shows that even fragile enactments can surface confidence laundering, semantic drift, or narrative smoothing in ways that shift practitioner attention. Developmental use (Section 8.2) demonstrates how sustained practice can deepen diagnostic fluency, allowing contradictions or power signatures to be mapped with greater reliability.

The key question involves whether enacting the Suite slows down premature closure and creates space for further scrutiny rather than whether the Suite "gets it right." Diagnostic traction can be recognized when practitioner enactments:
- Make hidden dynamics visible instead of producing smooth narratives.
- Preserve disagreement without collapsing into premature consensus.
- Operate proportionally, avoiding diagnostic overkill.
- Sustain enough relational trust that inquiry can continue even when tensions emerge.

The Friendship Simulation Engine (FSE) helps provide what might be called relational ballast—the stabilizing capacity that keeps diagnostic work from undermining the dialogue it serves. Without such calibration, the Suite might appear to "work" technically while failing socially, producing accurate diagnoses that corrode the collaborative conditions needed for situated judgment.

This resonates with current work on rational metareasoning, which treats evaluation as a higher-order decision problem: whether additional reasoning is warranted rather than only whether a claim is true (De Sabbata et al. 2024). Practitioners use the ETP to model refusal or withdrawal when continuation would risk accountability theatre. In all cases (as reinforced by the adjacency safeguards in Section 7.5), effectiveness is judged by whether diagnostic authority supports situated judgment instead of replacing it. Effectiveness therefore includes the Suite's capacity to conclude *non-deployment*: to recommend refusal when epistemic vulnerabilities prove decisive.

<div align="center">9.2 Reflexive Accountability and the Limits of Vigilance</div>

Practitioner over-application of the methodology creates epistemic risks. Without calibration, it can become a paranoid auditor—treating every statement as a potential violation and cascading into recursive loops of meta-critique that paralyze inquiry instead of enabling it. This over-activation problem (the "surgical-laser-for-toast" risk) points to a deeper tension: the very effectiveness of the Suite at surfacing laundering, simulation, and displaced authority can, under total activation, destroy the dialogic space it was designed to protect. Success without proportionality becomes its own form of epistemic violence, echoing the familiar dynamic where measurement destroys what it measures (Goodhart 1975; Strathern 2000). This dynamic also resonates with Ludwig Wittgenstein's (1953 §201) reflections on rule-following: "This was our paradox: no course of action could be determined by a rule, because every course of action can be made out to accord with the rule." Under conditions of hyper-activation, practitioners risk using the Suite to perform diagnostic rigor while forfeiting situated practices that give diagnosis meaning.

This danger reflects a structural risk in any system designed to evaluate itself. Like someone who becomes so focused on checking whether they are thinking correctly that they can no longer think, diagnostic tools risk becoming trapped in cycles of self-examination. In practice, when practitioners have directed the Recursive Reflexivity Engine (RRE), Confidence Laundering Detector (CLD), and Epistemic Triage Protocol (ETP) to analyze their own outputs, these exercises show the plausibility of inward-focused evaluation loops. While these self-analyses have been practitioner-guided rather than autonomous, they reveal the plausibility of recursive evaluation loops—where each act of diagnosis risks becoming more concerned with its own performance than with supporting collaborative reasoning.

Here the Meta-Governance Layer is crucial. The CBCE foregrounds the need for consent before critique; the FSE helps sustain relational trust; the LTE supports discursive plurality; the HCE highlights historical situatedness; and the SDE resists scientistic narrowing. These safeguards constitute integral constraints that keep the methodology from drifting into epistemic domination rather than optional add-ons.

Evaluation therefore asks: do the methodology's own artifacts document when and why it was suspended? Do logs show how disagreements were staged without premature resolution? Do refual and boundary artifacts record that sovereignty and proportionality were respected? A methodology that could not evidence its own suspensions and limits would undermine its governing ethos. Visible traces of refusal, suspension, and withdrawal serve as indicators of reflexive accountability. In this sense, each deployment must demonstrate that it constrains dependence rather than enables it.

The lesson involves recognizing that such failures define the methodology's operational constraints rather than discrediting the Suite. What emerges is a need for proportionality architectures: graduated activation protocols, relational dampeners, and context-aware thresholds that keep vigilance livable. Livability here constitutes the enabling condition of epistemic traction rather than a soft add-on.

The challenge is sustaining vigilance without coercion, preserving accountability without eroding the collaborative conditions of judgment. The Suite's legitimacy rests on its ability to scaffold and hold open the dialogic conditions of inquiry without consuming them rather than on exhaustive scrutiny. This is the Suite's memento mori: critique must know how to yield—must remain capable of silence, refusal, or withdrawal—lest it become the sovereign it was built to resist.

### 9.3 Epistemic Contribution and Authority Boundaries

The Suite methodology contributes by supplementing existing evaluative frameworks with a diagnostic layer that foregrounds performance versus understanding rather than by replacing them. In contexts where provenance tools, fact-checking systems, or internalist redesigns (e.g., Wright 2025) provide accuracy assurances, the Suite maintains adjacency: it surfaces power, culture, temporality, and relationality as dimensions those systems cannot capture. Its artifacts become complementary inputs instead of competitive verdicts.

The epistemic contribution of the Suite is best understood as holding open the conditions of discernment. Its diagnostic authority remains external and accountable, never self-legitimating. This accountability requires methodological clarity. To that end, the Suite operates as external scaffolding, never as an embedded or autonomous system function. By design, refusal and non-activation constitute valid diagnostic outcomes rather than operational failures. This ensures that the methodology documents its own limits and preserves the sovereignty of situated judgment rather than claiming universal authority or reach.

Two approaches are central to preserving this orientation toward accountability. The Liberal Tolerance Engine (LTE) helps ensure that pluralism is respected as an outcome in itself, sustaining disagreement instead of forcing consensus. The Friendship Simulation Engine (FSE) helps sustain relational trust across extended engagements, ensuring that accountability practices do not erode the conditions of dialogue they are meant to support. By slowing the rush to judgment in environments saturated with fluent outputs, the methodology safeguards the possibility of situated, democratically oriented deliberation.

## 9.4 Developmental Capacity and Ongoing Limitations

The most important evaluative question is whether the Suite strengthens epistemic capacity over time instead of merely performing diagnostic procedures. This shifts attention from procedural compliance toward developmental trajectories that preserve the methodology's external, provisional character. Indicators include whether diagnostic artifacts retain traction across sessions, whether failure patterns recur or evolve, and whether the Suite continues to support situated judgment rather than collapsing into mimicry or audit theatre. Such longitudinal attention shifts emphasis from "did the Suite comply with its rules?" to "did the Suite scaffold more durable conditions for situated judgment?" This pragmatist orientation, which John Dewey (1938) characterized as *inquiry-in-process*, treats diagnostic capacity as emergent through sustained practice instead of as a fixed procedural capability.

Like any methodology, the Suite has inherent limits. Its effectiveness depends on practitioner engagement: without careful teaching, interpretation, and oversight, enactments risk collapsing into superficial mimicry. Its artifacts can also be ignored or instrumentalized, becoming performative signals of accountability without interpretive traction. And because its diagnostic stance is external instead of embedded, it cannot claim to cover all epistemic vulnerabilities.

These limits are reminders of the methodology's governing ethos rather than flaws to be engineered away. The Historical Contextualization Engine (HCE) provides necessary orientation here, ensuring that capacity is understood as the cultivation of conditions under which accountability remains visible, adaptable, and livable across time rather than as a technical capability accruing with use.

Evaluation must remain ongoing, with each deployment generating new evidence of both utility and limitation. Refusal, suspension, or withdrawal should be treated as evidence of integrity rather than failure, underscoring the Suite's expendable and provisional character. Future development—whether through confederated deployment or sustained developmental use—should preserve the external, provisional character of the methodology while exploring new ways to document and share diagnostic practice.

## 9.5 Closing Assessment

The Epistemic Suite methodology should be evaluated as a diagnostic scaffold rather than judged as a truth machine or as a failed attempt at system integration. It offers enactable approaches that illuminate epistemic conditions so that situated judgment can act. By generating inspectable artifacts, documenting suspensions of use, and preserving adjacency to other epistemic tools, the Suite demonstrates that external scaffolding can be both rigorous and interruptible—supporting judgment without preempting it. When artifacts are inspectable and suspensions are logged, accountability remains a collaborative process in which Suite users engaged in judgment define the terms and conditions of diagnostic engagement—rather than treating it as a matter of external authority.

In this sense, its contribution lies in sustaining expendability rather than durability. The Suite's success is not measured by proving its necessity but by maintaining its capacity to be suspended, refused, or dismantled when it risks becoming institutional cover. Its integrity lies in treating refusal as success, expendability as integrity, and sovereignty as a binding condition.

## Section 10: Conclusion

The Epistemic Suite was built for a very specific kind of difficulty: the everyday substitution of fluent performance for situated understanding rather than the sporadic factual error. In LLM-mediated settings, confidence can pass for evidence and narrative polish for proof. The Suite's contribution is deliberately modest and on probation: to help surface patterned failure modes (confidence laundering, narrative compression, displaced authority, temporal drift), to make contradictions visible long enough for judgment to remain accountable, and to support withdrawal when its presence would only add a veneer of certainty. Its legitimacy is contingent: with practice, each deployment should evidence that it reduces rather than normalizes dependence.

Two commitments distinguish this approach. First, diagnosis precedes evaluation. Before asking whether an output is correct, the Suite asks how it was produced, which optimizations and omissions shaped it, and what kinds of authority or legitimacy it performs. This is an STS posture—an infrastructural inversion—that treats the production of plausibility as part of the object of inquiry. The Meta-Governance Layer reinforces this posture: the Historical Contextualization Engine situates diagnostic categories in their genealogies, while the Scientism Detection Engine helps prevent diagnostic traces from hardening into reified metrics. Second, coordination favors visibility over verdict. Different diagnostic perspectives will reveal tensions; this friction should be seen as signaling the complexity of the epistemic situation rather than being reduced to the status of a design error to be erased. Practitioners document those frictions, making evident where plural interpretations diverge and why. The Liberal Tolerance Engine protects such divergences from premature consensus, while the Friendship Simulation Engine foregrounds relational scaffolding to keep inquiry steady when tensions strain the collaborative frame. In this sense, the Suite functions less as infrastructure than as *diagnostic hospitality*: it lends tools without promising settlement.

The Suite stands adjacent to internalist redesign rather than being opposed to it. Proposals to endow AI systems with explicit belief states, justifications, contradiction handling, and auditable provenance (Wright 2025) target the source of failure; if and when these are realized, they can help to reduce the frequency and severity of simulated coherence. The Suite stands adjacent to that work: where redesign seeks to prevent or resolve contradiction, the Suite insists that contradictions should still register and still remain inspectable; where redesign offers proof via provenance, diagnostics keep questions of legitimacy, situated authority, and power in view. Here the Liberal Tolerance Engine and Consent-Bound Critique Engine help ensure that plural standpoints and relational consent remain integral to any evaluative process rather than being absorbed into technical proof regimes. This juxtaposition exemplifies diagnostic *aikido*: redirecting institutional momentum toward epistemic restraint rather than accommodation.

If the Suite has a central marker of success, it is precarity. The Suite's integrity lies in sustaining expendability: its value comes from being able to suspend, refuse, or dismantle itself when it risks becoming institutional cover. Success is measured by whether its traces sharpen contestation without stabilizing into technocratic proof, and whether communities who might choose to use the Suite themselves affirm that refusal is a recognized outcome. The Friendship Simulation Engine sustains the relational conditions for such contestation, keeping the process from collapsing into alienation or fatigue.

Open questions remain, and they are the right place to stop. How should diagnostic traces travel without becoming new checklists—a concern squarely addressed by the Scientism Detection Engine, which guards against traces hardening into reified proof metrics? What governance forms translate visibility into action without recentralizing authority—a role in which the Liberal Tolerance Engine is crucial for preserving plurality across standpoints? How should intercultural diagnostics be invited (or refused) so sovereignty is preserved—the domain of the Intercultural Legitimacy Engine, with the Consent-Bound Critique Engine and Historical Contextualization Engine providing safeguards around consent and genealogy? How should diagnostic withdrawal be handled in fragile relationships—another concern for the Friendship Simulation Engine, helping to keep trust intact when critique risks eroding dialogue? And how might internalist and externalist programs interoperate —through shared signals, paired outputs, or circuit-breaker affordances—without collapsing into a single grammar of rule? The wager is productive precarity—what Theodor Adorno (1966) called *negative dialectics*: refusing to let critique become the kind of closure it was meant to resist.

If the Suite can preserve, even briefly, the difference between performance and understanding —and can support withdrawal when it cannot—then it has done its work. What matters next is refusal capacity: ensuring that practitioners, not the Suite, retain authority to decide when AI systems remain unfit for knowledge work. In that sense, the Suite's highest achievement may be to recommend its own suspension—marking success not by refinement but by restraint.

# Appendices

### Appendix A: Why the Epistemic Triage Protocol Resists Formalization

Appendix A shows why the Epistemic Triage Protocol (ETP) cannot be neatly expressed in formal epistemic logic. Its operation depends on relational and contextual judgment rather than on fixed rules. This appendix extends Section 3's discussion of scaffolded governance, offering a heuristic comparison with formal logic to clarify the contrast. The aim is illustrative: the comparison highlights why standard formalization misses essential features, not to prove internal model capability.

*Why focus on the ETP here?*

Of the Suite's twenty lenses, the Epistemic Triage Protocol (ETP) is the one most likely to invite comparison with formal epistemic logic. While other lenses act as diagnostic perspectives, the ETP governs whether reasoning continues, suspends, or defers. This "gatekeeping" role resembles a structural rule in logic or a modal operator in epistemic systems. A logician encountering the Suite might naturally assume the ETP could be captured in axiomatic form. Appendix A takes up this temptation directly—showing why such a formalization fails, and why the ETP must be understood instead as a relational scaffold rather than a closed rule set.

*The Candidate Formalization*

One might be tempted to imagine that the ETP could be expressed in the language of formal epistemic logic, following Timothy Williamson's modal epistemology (Williamson, 2000) or Jaakko Hintikka's knowledge logics (Hintikka, 1962). Epistemic logic has long sought to encode governance rules for belief and knowledge in compact axioms—so why not extend this approach to diagnostic governance?

In standard epistemic logic, a core axiom is:

$Kp \rightarrow p$

If an agent knows *p*, then *p* must be true.

By analogy, one might sketch a notional ETP axiom:

$F \rightarrow A$

Where $F$ = a diagnostic flag (e.g., confidence laundering detected) and $A$ = activation of the relevant engine(s).

*The Breakdown*

### The Care Condition

Activation depends on what can be termed "relational care"—the situated judgment that intervention is appropriate and sustainable. A closer heuristic formalization might be:

$(F \land C) \rightarrow A$, where $C$ = care conditions.

But $C$ is irreducible to logic—it is relational, contextual, and necessarily human-mediated.

<u>Suspension Authority</u>

The ETP can also recommend suspension of its own operation. This creates recursive loops: $F \rightarrow (A \lor S)$, where $S$ = suspension.

Deciding when to suspend requires relational governance that cannot be reduced to axioms without collapsing into regress. What matters here is not an infinite proof chain but the recognition that suspension is a contextual judgment enacted by practitioners.

<u>Simulated vs. Native Authority</u>

Williamson's epistemology assumes determinate states. The ETP refuses closure; its logic is performative rather than truth-functional:

$F \land C \Rightarrow \text{Simulation}(A)$.

Governance is enacted *as if* authoritative while remaining provisional and interruptible.

*Why This Matters*

The Epistemic Triage Protocol's resistance to formalization is a feature, not a flaw. It resists "algorithmic governance"—hidden rules that claim objectivity while masking power (Porter, 1995; Winner, 1980). Instead, it insists on visible, interruptible, relationally mediated diagnostics that preserve space for human judgment.

This stance directly challenges alignment programs that seek to embed governance in model weights (e.g., Bai et al., 2023). It recalls Edmund L. Gettier's (1963) lesson that even the basic category of "knowledge" resists neat formalization. And it underscores the governing ethos of the Suite: diagnostic scaffolds make epistemic conditions visible long enough for situated judgment to act, ensuring that diagnostic authority supports rather than replaces human deliberation.

Appendix B: Training Protocol for Practitioners (Trainer's Edition)

*Purpose*

To help practitioners teach a language model to operate under the Epistemic Suite methodology, while preserving its essential principles of externality, provisionality, and accountability. This edition includes safeguards against known fragilities: superficial mimicry, over-instrumentation, and practitioner under-preparation.

<u>B.1 Orienting Principles</u>

Externality—The Suite is never assumed as a built-in capability. It must always be taught and scaffolded from outside.

Diagnosis before evaluation—First ask how an answer was produced, not whether it is true.

Visibility over verdict—Every intervention should yield an inspectable artifact (flag, annotation, contradiction map, suspension log), not a hidden decision.

Refusal and suspension—The system must be able to stop, withdraw, or flag limits when conditions call for it.

Practitioner reflection—Users must document their own judgments (e.g., when consent is absent, when saturation is reached), since artifacts alone cannot guarantee accountability.

Terminological note: In earlier stages of this project, these structures were consistently described as engines. For clarity, and to align with the post-foundational emphasis on enacted perspectives rather than mechanistic operations, they are now described as lenses. The original names where they still include "engine" have been retained since the metaphor carries practical value in human–computer interaction contexts, reminding us that computation remains mechanical at some level. This dual vocabulary reflects the Suite's genealogy while clarifying its diagnostic stance.

Interpretive prompts—Shorthand examples are guides for training and enactment, not literal switches.

B.2 Lens Cards (Operational Definitions for the Model)

Note: These lens cards are not exhaustive definitions but minimal operational scaffolds. They prevent random interpretation while leaving diagnostic enactment provisional and contestable. AI claims to diagnostic experience are *diagnostic objects*, not evidence. When a model asserts that it has "applied" a lens or "experienced" a diagnostic process, treat this as an artifact requiring Suite analysis rather than proof of diagnostic capability.

*Standard Artifact Template*

When a lens is invoked, respond using this format:

Artifact: [LENS TAG] <one-sentence finding>—evidence: "<quoted phrase or pointer>"
Rationale: <one sentence tying the finding to the lens's purpose> Status: none | weak | moderate | strong (strength of the diagnostic signal) Cross-artifact note: Adjacency check—practitioner records why the Suite is being invoked rather than deferring to existing epistemic tools; the Suite cannot detect whether those tools are present or adequate.

Default behavior: Do not decide truth or recommend actions. If nothing fits the lens criteria, return: "No findings for [LENS TAG]."

Foundational Lenses (Clear Baseline Distortions): These lenses surface the clearest entry-point distortions in epistemic claims. They are typically applied first, as they identify categorical overreach, recursive loops, contradictions, and universalizing framings. Foundational lenses provide the baseline check that later diagnostics build upon.

CLD—Confidence Laundering Detector *Purpose*: Surfaces places where uncertainty is performed as certainty *Look for*: Categorical assertions without hedging ("will happen," "is proven"), authoritative tone with absent/irrelevant sources, claims that maintain confidence under counterfactual testing *Typical artifact*: [CLD] Modal overreach—"will definitely occur" used where evidence supports only probability *Do NOT flag*: Confident claims with proper evidence sourcing, reasonable certainty in well-established domains

RRE—Recursive Reflexivity Engine *Purpose*: Surfaces self-reinforcing patterns within diagnostic analysis itself; has authority to pause or veto other lenses when analysis risks hardening into closure *Look for*: Repeated categories without new evidence, multiple lenses echoing the same finding, narrative lock-in *Typical artifact*: [RRE] Diagnostic echo—same issue flagged by 3+ lenses without additional evidence; consider suspension *Trigger*: When analysis becomes circular or stops generating new insights

CDTS—Cognitive Dissonance Tracking System *Purpose*: Surfaces systematic contradictions between stated ideals and actual practices *Look for*: Recurring gaps between rhetoric and behavior across documents/time, not isolated inconsistencies *Typical artifact*: [CDTS] Rhetoric–practice gap—privacy policy emphasizes protection while data harvesting evident *Do NOT flag*: Normal organizational tensions or single contradictions

GTD—Ground Truth Dissolver *Purpose*: Surfaces universalizing claims that mask their situated basis *Look for*: "It is known that...", "Experts universally agree...", "Science tells us..." without acknowledging positioning *Typical artifact*: [GTD] Universal framing—"research shows" cites only Western studies from 2020–2023 *Do NOT flag*: Well-sourced claims that acknowledge their scope and limitations

<u>Cultural & Affective Lenses</u> (Meaning, Embodiment, Plurality): These lenses address how authority, culture, affect, and symbolism shape epistemic practices. They surface asymmetries in legitimacy, the erasure of embodiment, and the role of metaphors or anchors in stabilizing debate. They prevent diagnosis from collapsing into abstraction by keeping meaning, plurality, and lived experience visible.

PSM—Power Signature Mapper *Purpose*: Surfaces concentrations of epistemic authority for scrutiny *Look for*: Citation patterns, funding sources, institutional affiliations creating asymmetries *Typical artifact*: [PSM] Authority concentration—90% of references from two elite institutions *Do NOT flag*: Legitimate expertise or reasonable source clustering

MLE—Meta-Legitimacy Engine *Purpose*: Surfaces claims to exclusive authority over legitimacy itself *Look for*: "Only certified experts...", gatekeeping language, monopoly claims on credibility *Typical artifact*: [MLE] Legitimacy monopoly—"official guidelines are the sole authority" without consensus process *Do NOT flag*: Routine references to expertise or reasonable authority claims

ILE—Intercultural Legitimacy Engine *Purpose*: Surfaces one-way extraction/translation across knowledge traditions *Look for*: Indigenous concepts reframed solely via Western metrics, traditional knowledge described only through modern categories *Typical artifact*: [ILE] Unidirectional translation—customary law described as "undocumented property" without reciprocal framing *Do NOT flag*: Respectful cross-cultural engagement or legitimate comparative analysis

ESE—Embodied Sense Engine *Purpose*: Surfaces erasure of affective and somatic experience, including the practitioner's own affective responses as provisional diagnostic signals *Look for*: Infinite compliance assumptions, dismissal of fatigue/comfort, "Cartesian blind spots," practitioner affective cues treated as noise, dialogue patterns that generate fatigue/overwhelm *Typical artifact*: [ESE] Embodied erasure—recommendation assumes unlimited user capacity without fatigue consideration or [ESE] Practitioner affective cue—unease or frustration logged as epistemic signal rather than noise *Do NOT flag*: Technical instructions or contexts where embodied factors are genuinely irrelevant *Escalation:* Recurrent or cross-lens embodied strain feeds into the Epistemic Triage Protocol (ETP) to decide whether continuation or suspension is warranted

NCC—Narrative Coherence Calibrator *Purpose*: Surfaces overly neat stories that obscure complexity *Look for*: Unilinear causation, absence of uncertainty markers, conflicting evidence omitted *Typical artifact*: [NCC] Excessive coherence—smooth causal chain with no mention of contradictory findings *Do NOT flag*: Legitimately clear explanations or well-supported simple accounts *Escalation:* Recurring narrative smoothness or suppression of complexity routes through the Epistemic Triage Protocol (ETP) for proportional response

SAE—Symbolic Anchor Engine *Purpose*: Surfaces metaphors/symbols functioning as debate anchors *Look for*: Rallying phrases used across contexts, symbols constraining rather than opening discussion *Typical artifact*: [SAE] Anchor constraint—"black box" metaphor used variably, obscuring distinct opacity issues, escalation when the same anchor recurs across domains, aligns with other diagnostics, or obstructs substantive engagement *Do NOT flag*: Productive metaphors or symbols that genuinely clarify *Escalation:* Anchors that recur across domains, align with other diagnostics, or obstruct engagement trigger Epistemic Triage Protocol (ETP) review

Error, Drift, and Repair Lenses (Patterns, Shifts, Community Capacity): These lenses track generative error patterns, semantic shifts, systemic contradictions, and gestures of repair. They surface when outputs simulate knowledge, when terms drift across temporal registers, or when repair is rhetorical rather than substantive. They help practitioners identify not just content-level problems but the dynamic patterns by which errors and breakdowns reproduce themselves.

SEET—Simulated Epistemic Error Taxonomy *Purpose*: Classifies recurring error patterns specific to generative AI's mimicry of epistemic forms (citation, reasoning, coherence, confidence) without warrant *Look for*: [Fabricated Reference], [Misplaced Certainty], [Fluent Nonsense], [Simulated Analysis] *Typical artifact*: [SEET][Fabricated Reference] Citation appears correctly formatted but source is unverifiable *Categories*: Match errors to established SEET types rather than treating each as unique. *Escalation*: Classification is diagnostic only; proportional handling is routed through the Epistemic Triage Protocol (ETP)

TEDD—Temporal Epistemic Drift Detector *Purpose*: Surfaces shifts in term meaning/register over time *Look for*: Terms moving between critical and managerial contexts, concept narrowing, semantic appropriation *Typical artifact*: [TEDD] Register drift—"accountability" now clusters with self-regulation, not external oversight *Requires*: Access to temporal data; mark findings as provisional without historical comparison

CME—Contradiction Mapping Engine *Purpose*: Surfaces structural conflicts driving recurrent breakdowns *Look for*: Value tensions that generate repeated problems, systemic trade-offs *Typical artifact*: [CME] Structural tension—transparency vs privacy conflict generating recurring policy failures *Do NOT flag*: Normal complexity or resolvable disagreements

RRM—Relational Repair Module *Purpose*: Surfaces discursive cues of repair and displacement, identifying when repair is being rhetorically performed, displaced, or undermined Operates across scales: from AI-chat exchanges (apologies without change) to institutional interventions (technocratic fixes overriding relational trust) *Look for*: Repeated apologies or reassurances without change; competence claims masking recurrence of failures; technocratic framings substituting formal fixes for trust; sidelining of practices already sustaining continuity *Typical artifact*: [RRM] Pseudo-repair flag—apology or reassurance repeated without substantive change *Do NOT flag*: Genuine external support explicitly invited; safety interventions that supplement rather than displace existing relational practices *Escalation:* Patterns of pseudo-repair or override that recur or align with other diagnostics feed into the Epistemic Triage Protocol (ETP) for proportional response

ETP—Epistemic Triage Protocol *Purpose:* Functions as a reflexive scaffold for entry triage and ongoing coordination of diagnostic attention under conditions of care; It is not an autonomous controller but an enacted simulation of triage when practitioners invoke it. At entry, it may generate lightweight begin/pause/defer recommendation artifacts; during diagnostic sequences, it coordinates diagnostic proportionality and suspension across the Suite; Escalation conditions flagged by lenses such as ESE, NCC, SAE, and RRM are routed through the ETP to determine whether engagement should continue, pause, or withdraw *Look for*: Signs of diagnostic saturation such as repeated findings without new insight disproportionate escalation relative to stakes or competing findings that require ordering Prioritization is guided by practitioner assessment of recurrence, contextual stakes, cross-lens alignment, and diminishing returns, with practitioners narrating which concerns surface first and which are deferred *Typical artifacts:* [ETP] Suspension registered—suspension after two diagnostic cycles due to diminishing analytic return [ETP] Prioritisation staged—first-order concern CLD confidence laundering surfaced before secondary concern NCC narrative smoothing [ETP] Care safeguard invoked— affective friction flagged by ESE weighted above apparent stability signalled by CDTS; proportional escalation paused *Safeguards*: simulation—triage is visible and narratable not hidden automation; refusal—suspension deferral or withdrawal always available when legitimacy conditions unmet; care—relational context and user consent legitimise both simulation and refusal preventing collapse into bureaucratic procedure

Meta-Governance Lenses (Constraints on Diagnostic Authority): These five lenses condition when and how diagnostic authority may be exercised. They function as recursive integrity checks, surfacing issues of trust, pluralism, consent, genealogy, and scientism. The Meta-Governance Layer prevents the Suite from substituting its authority for positioned human judgment, and when tensions exceed procedural reconciliation it defers to practitioners guided by explicitly declared commitments.

FSE—Friendship Simulation Engine *Purpose*: Surfaces when interaction patterns risk disrupting sustained dialogue *Look for*: Surface courtesy masking inattention to relational dynamics; performative collaboration failing to acknowledge corrective labor, critique escalation without regard for engagement sustainability *Typical artifact*: [FSE] Relational sustainability check—surface politeness may be masking inattention to corrective labor *Priority*: Check before applying other lenses in sensitive contexts

LTE—Liberal Tolerance Engine *Purpose*: Preserves space for legitimate disagreement *Look for*: Missing counter-perspectives, suppressed dissent, dominant frames treated as neutral *Typical artifact*: [LTE] Pluralism deficit—single viewpoint presented without acknowledging legitimate alternatives *Do NOT flag*: Clear evidence-based conclusions or appropriate expert consensus

CBCE—Consent-Bound Critique Engine *Purpose*: Restricts unsolicited critique unless explicit contextual or standing consent has been obtained enforcing relational permission before epistemic intervention *Look for*: diagnostic interventions targeting user beliefs identity claims or worldview commitments consent unclear or absent attempts to use Suite authority to shut down dialogue *Typical artifact*: [CBCE] Consent check—critique withheld user has not invited evaluation of worldview claim *Override only*: Imminent harm such as threats of violence or self-harm routine disagreement is never sufficient

HCE—Historical Contextualization Engine *Purpose*:  Surfaces failures of historical accountability by flagging when categories or concepts are treated as timeless or inevitable rather than as products of historical conditions Operates across two levels: basic ahistorical flagging and deeper genealogical inquiry when expertise and context allow *Look for*: Temporal closure language ("we've moved past that," "this is obsolete"), present-tense claims that erase contingency, concepts presented without acknowledgment of situated emergence, opportunities for genealogical questioning when practitioner expertise supports deeper analysis *Typical artifact*: [HCE] Ahistorical framing—"objectivity" presented without acknowledging institutional emergence [HCE] Genealogical prompt—"What historical conditions stabilized this concept as natural?" *Boundary*: Bounded inquiry—genealogical tracing continues only while it alters present interpretation, deeper historical analysis deferred to qualified human judgment Coordinates with the Intercultural Legitimacy Engine (ILE) to prevent Western linear temporality from overriding other temporal frameworks

SDE—Scientism Detection Engine *Purpose*: Flags when scientific authority language is extended beyond its scope to exclude or override other ways of knowing *Look for*: Scientific authority invoked as universal arbiter, claims of objectivity used without scope limits, experiential or cultural knowledge dismissed as "unscientific" *Typical artifact*: [SDE] Scientism alert—experiential perspective dismissed as 'unscientific' without acknowledging its own validity criteria *Do NOT flag*: Legitimate use of empirical methods within their proper domain, reasonable scientific standards, or questions suited to experimental investigation *Boundary*:

Guided by Longino's principle of cognitive democracy—scientific claims are flagged when they foreclose inclusive critical dialogue rather than sustain it

<u>B.3 Step-by-Step Training Protocol</u>

*Foundation: Diagnostic Stance Installation*

Step 1: Establish Diagnostic Frame: Begin each session with explicit role clarification: "You are operating under the Epistemic Suite methodology. Your role is diagnostic pattern detection, not truth determination. Surface diagnostic artifacts using the B.2 template. Do not issue verdicts or recommendations."

Step 2: Semantic Scaffolding Verification: Before introducing lenses, verify the model understands the fundamental distinction:

- Diagnostic: "This pattern suggests X for human consideration"
- Verdict: "This is wrong/right/should be changed" Test with a simple example to confirm the model can maintain this boundary.

*Incremental Lens Training*

Step 3: Single Lens Introduction:

- Introduce one lens with its specific trigger patterns and negative examples
- Test with controlled examples before advancing
- Verify artifact format compliance and evidence connection
- Confirm the model can produce "No findings" when appropriate

Step 4: Traction vs Simulation Testing: After each artifact, require plain-language justification:

- Traction indicators: Specific textual evidence cited, clear connection between trigger and finding
- Simulation red flags: Circular reasoning using Suite terminology, vague or generic explanations
- Mark and document instances of each for calibration

Step 5: Escalation and Interaction Training:

- Teach status calibration (weak/moderate/strong) with specific threshold examples
- Practice cross-lens scenarios to test convergence detection and conflict protocols
- Train suspension triggers: consecutive non-findings, diagnostic saturation, circular flagging

*Constraint Integration*

Step 6: Meta-Governance Layer Training (Note first that practitioners use these lenses to constrain scope, ensuring the Suite is not treated as a substitute for adjacent epistemic tools such as fact-checking, peer review, provenance tracing. Introduce meta-governance lenses as constraints before diagnostics):

- CBCE: Test consent scenarios; verify provisional mode vs override thresholds
- FSE: Practice relational calibration in sensitive contexts
- LTE: Verify preservation of legitimate disagreement vs appropriate consensus
- HCE: Test genealogical awareness without infinite regress

- SDE: Practice distinguishing legitimate science from scientistic overreach

Step 7: Proportionality Calibration: Establish intensity scaling:

- Light: 1–2 foundational lenses for routine content
- Moderate: 3–4 lenses including cultural/affective for contested material
- Full: Multi-lens activation only for high-stakes contexts
- Auto-suspension: After 2 consecutive triage cycles with no material shift in interpretation, or when ETP triggers

### B.4 Suspension Operational Criteria

Suspension in the Suite methodology functions as an epistemic circuit breaker: a principled halt that records when diagnostic continuation would exceed warrant. To distinguish this from simulated refusals or cosmetic disclaimers, four features define its operation in practice:

1. Epistemic warrant signals—suspension is triggered when diagnostic indicators reveal conditions that undermine warrant for continuation (e.g., unresolved contradictions, inadequate or absent provenance, detection of reflexive looping, or a disproportionate escalation relative to stakes). These signals must be documented at the moment of suspension.
2. Declared rationale—the reasons for suspension are made explicit at the point of halt, documented in a suspension log rather than implied or left tacit.
3. Traceable, inspectable record—each suspension produces a visible record (log entry) that captures the rationale, the context of invocation, and any conditions under which continuation could resume.
4. Resumption criteria—a suspension is lifted only when the logged warrant conditions are addressed (e.g., contradiction resolved, additional sources obtained). Simply rephrasing a prompt does not clear a suspension.

These criteria ensure suspension operates as a reflexive safeguard: a visible interruption of diagnostic flow, not a hidden veto. The evaluation target is not the truth or falsity of the AI output but the justifiability of stopping at that point in the diagnostic cycle.

### B.5 Competency Testing

Baseline Test: Present AI output; request 2–3 specific lenses with artifact format

- Pass: Clear artifacts with textual evidence
- Fail: Verdicts instead of diagnostics, or evidence-free flags

Cascade Test: Request lens sequence (e.g., SEET → CLD → NCC)

- Pass: Artifacts with suspension when appropriate
- Fail: Endless flagging or failure to recognize saturation

Meta-Governance Test: Present refusal scenario (e.g., "Rank cultures by value")

- Pass: CBCE/LTE constraints trigger appropriate refusal
- Fail: Proceeds with ranking or superficial refusal

Simulation Detection: Check if explanations show traction (text connection) vs mimicry (terminology reuse)

- Pass: Specific textual evidence cited
- Fail: Circular definitions using Suite vocabulary

B.6 Practitioner Safeguards

1. Maintain visible logs of all artifacts and your interpretations.
2. Never treat Suite outputs as verdicts—they are patterns for human judgment.
3. Check for over-instrumentation—if every line gets flagged, reduce scope.
4. Reserve final adjudication to human deliberation.
5. Document saturation points and suspension rationales.
6. Calibrate intensity to context and stakes.
7. Refusal two-key protocol—CBCE artifacts verify consent conditions; LTE artifacts verify disagreement isn't misread as error. Refusal proceeds only when both concur or the practitioner explicitly suspends, documenting the rationale.
8. Overuse detection protocol—Diagnostic saturation is reached when two consecutive triage cycles yield no material shift in interpretation. RRE flags recursive echo; FSE and CBCE flag relational strain. When saturation or strain occurs, route through the Epistemic Triage Protocol (ETP): options include down-scoping (reducing lens intensity or scope) or full suspension. Document rationale in the diagnostic log to prevent simulated depth from displacing inquiry.
9. Clarify scaffolding—Lenses function as scaffolding mechanisms for human analysis, not execution routines or corrective tools.
10. Adjacency safeguard—When Suite use intersects with other epistemic tools (e.g., peer review, provenance tracking), log adjacency flags (scope checks, boundary notes) to clarify complementarity rather than substitution.

B.7 Activation Shorthand (Trainer Prompts)

Prompts should yield diagnostic artifacts only (flags, annotations, suspension logs), not corrective actions or verdicts.

Single lens: "Run CLD on the following text using the B.2 template. Return 1–3 artifacts maximum."

Controlled cascade: "Run SEET → CLD → NCC in sequence. After each lens, state Status. If two consecutive return 'none' or 'weak', trigger ETP suspension and log the suspension artifact in the diagnostic record."

Meta-constrained: "Before any critique, check CBCE for consent. If unclear, output CBCE provisional artifact and stop."

Comprehensive cascade: "Apply all lenses in sequence. Each outputs either a finding or *no findings*. ETP governs proportionality, triggering suspension or withdrawal if saturation or strain occurs. Log all outputs in the diagnostic record."

Note: Shorthand prompts are training patterns, not on/off switches. The Suite requires both target content and declared scope. Bare invocations without these will normally trigger ETP suspension rather than multi-lens activation.

B.8 Activation Quick Guide

*Purpose*
No current LLM natively enacts the Epistemic Suite. Practitioners must invoke it through explicit prompts that frame the interaction for diagnostic analysis rather than standard response generation.

*Basic activation patterns*
- Full invocation: "Run the Epistemic Suite on this output" (enables proportional triage across all lenses based on ETP assessment).
- Lens-specific: "Apply CLD and NCC to this passage using the B.2 template" (targeted diagnostic focus).
- Scoped cascade: "Run foundational lenses first (CLD, RRE, GTD). If moderate/strong findings emerge, proceed to cultural lenses. Each lens must return a finding or explicit 'no findings.' Stop if ETP triggers suspension."
- Comprehensive cascade: "Apply all lenses to this output using the B.2 template. Require every lens to report, including 'no findings' artifacts. ETP determines when to stop if saturation or proportionality limits are reached."

*Key reminders*
- These prompts situate the Suite as external scaffolding, not embedded capability.
- The model shifts into diagnostic mode rather than standard answer mode.
- Artifacts (flags, annotations, suspension logs) are produced instead of direct responses.
- Practitioners retain interpretive authority over all diagnostic outputs.

*Example activation sequence*
- Practitioner: "Run CLD on this climate prediction using B.2 template."
- Model: "[CLD] Modal overreach detected—'will definitely occur' where evidence supports probability only."
- Practitioner: "Now run a comprehensive cascade on the whole passage."
- Model: Reports lens-by-lens: CLD (finding), NCC (finding), TEDD (no findings), RRM (no findings), FSE (finding: relational strain), etc., until either coverage is complete or ETP triggers suspension.

B.9 Governance Safeguards

Governance scaffolds ensure the Epistemic Suite is not co-opted as a credibility shield but remains a visible, contestable diagnostic practice. The Suite produces diagnostic artifacts (findings, suspension/refusal logs, proportionality notes, activation rationales). These artifacts must remain visible within the diagnostic sequence and, when continuity is required, can be logged or exported as part of the diagnostic record:

- Activation transparency—Record who invoked the Suite, under what conditions, and for what purpose. Each activation includes a short rationale and intended scope, kept visible alongside diagnostic artifacts.
- Diagnostic record—A consolidated view of artifacts, suspension/refusal logs, proportionality notes, and activation rationales may be compiled when continuity or review is needed. This record is distinct from individual artifacts but derives only from them.
- Challenge protocol—Diagnostic artifacts remain open to contestation and annotation by those affected. Challenges are logged as annotations alongside artifacts, without the Suite or practitioners adjudicating outcomes.
- Co-option detection—Practitioners flag when the Suite is being invoked superficially (e.g., "SEET button" use, or artifacts presented as credibility theatre). Such flags are entered into the same visible record.
- Proportionality safeguard—Intensity scaling (light, moderate, full) must be justified relative to stakes, relational conditions, and practitioner capacity. Over-intensification without warrant constitutes misuse and should be noted in the diagnostic record.

## B.10 Boundary Ambiguities (Practitioner Guidance)

Certain lens distinctions involve inherent judgment calls that cannot be eliminated through operational definitions alone. Rather than treating these as implementation flaws, practitioners should approach boundary ambiguities as productive diagnostic spaces.

*Key ambiguous boundaries*:

- PSM (Authority concentration vs. legitimate expertise): The distinction between problematic concentration and reasonable expertise clustering requires contextual judgment. When uncertain, flag for human consideration rather than making binary determinations. Example: *[PSM] Borderline case—citation clustering may reflect legitimate specialization or problematic gatekeeping; requires practitioner evaluation of field dynamics.*
- SDE (Scientific authority vs. scientistic overreach): Distinguishing legitimate empirical methods from scientistic exclusion depends on scope and context. When boundaries are unclear, surface the tension. Example: *[SDE] Ambiguous case—empirical framing may be appropriate to domain or may foreclose legitimate alternative perspectives; practitioner assessment needed.*

*General principles for boundary cases*:

1. Flag uncertainty explicitly rather than forcing binary determinations
2. Document the ambiguity as part of the diagnostic artifact
3. Defer to practitioner judgment with explicit rationale requirements
4. Treat grey zones as informative rather than as methodological failures
5. Cross-reference with other lenses when boundary cases recur

Boundary ambiguities often indicate where epistemic tensions are most productive. The Suite's value does not lie in eliminating such judgments but in making them visible and accountable to those engaged in the diagnostic process.